\documentclass[times,doublespace]{simauth}

\usepackage{moreverb}

\usepackage[colorlinks,bookmarksopen,bookmarksnumbered,citecolor=red,urlcolor=red]{hyperref}

\newcommand\BibTeX{{\rmfamily B\kern-.05em \textsc{i\kern-.025em b}\kern-.08em
T\kern-.1667em\lower.7ex\hbox{E}\kern-.125emX}}

\def\volumeyear{2015}

\usepackage{lscape,rotating}
\usepackage{float,epsfig}
\usepackage{subfigure}
\usepackage{setspace}
 \usepackage{amsmath,bm} 
\usepackage{amsfonts}
\usepackage{amssymb}

\newtheorem{example}{Example}

\newenvironment{proof}[1][Proof]{\begin{trivlist}
\item[\hskip \labelsep {\bfseries #1}]}{\end{trivlist}}

\newcommand{\av}{\bm{a}}

\newcommand{\xv}{\bm{x}}

\newcommand{\yv}{\bm{y}}

\newcommand{\zv}{\bm{z}}

\newcommand{\thetav}{\bm{\theta}}
\newcommand{\alphav}{\bm{\alpha}}
\newcommand{\betav}{\bm{\beta}}

\newcommand{\zerov}{\bm{0}}

\makeatletter
\long\def\@makecaption#1#2{%
  \vskip\abovecaptionskip
  \sbox\@tempboxa{#1: #2}%
  \ifdim \wd\@tempboxa >\hsize
    #1: #2\par
  \else
    \global \@minipagefalse
    \hb@xt@\hsize{\box\@tempboxa\hfil}%
  \fi
  \vskip\belowcaptionskip}
\makeatother

\begin{document}

\runninghead{Sample size for tests based on t distributions}

\title{A noniterative  sample size procedure for  tests based on  t distributions}

\author{Yongqiang Tang}

\address{Shire, 300 Shire Way, Lexington, MA 02421, USA}

\corraddr{E-mail: yongqiang\_tang@yahoo.com}

\begin{abstract}
A noniterative sample size procedure is proposed for  a general hypothesis test based on the t distribution by modifying and extending
 Guenther's (1981) approach  for the one sample and two sample t tests. 
The generalized procedure is employed to determine the sample 
size for treatment comparisons using the analysis of covariance (ANCOVA) and the mixed effects model 
for repeated measures (MMRM) in randomized clinical trials. 
The sample size is calculated by adding a few simple correction terms to the sample size  from the normal approximation
to account for the nonnormality of the t statistic and lower order variance terms, which are functions of the covariates  in the model.
 But it does not require specifying the covariate distribution. 
The noniterative procedure is suitable for superiority tests, noninferiority tests and a special case of the tests for equivalence or bioequivalence, and generally
yields the exact or nearly exact sample size estimate after rounding to an integer.
The method for calculating the exact   power of the two sample t test with unequal variance in superiority trials is extended to equivalence trials. 
We also derive accurate power  formulae for ANCOVA and MMRM, and the formula for ANCOVA is exact for normally distributed covariates.
Numerical examples demonstrate the  accuracy of the proposed methods  particularly in small samples. 
\end{abstract}

\keywords{ Analysis of covariance; Crossover trial; Exact power; Kenword-Roger variance; Mixed effects models for repeated measures;
Superiority, noninferiority, equivalence and bioequivalence  trials}

\maketitle

\section{Introduction}\label{sec:intro}
Many common tests for continuous outcomes are based on the t test statistics. Examples include the one sample t test, two sample t test, 
and tests associated with the analysis of covariance (ANCOVA) and 
linear mixed effects models for repeated measurement (MMRM). 
The  sample size  determination is critical to ensure the success of a clinical trial since an underpowered study has less chance to detect an important treatment effect, 
whereas  the samples that are too large may waste time and resources \cite{chow:2008}.
Sample size calculation for the t tests is usually based on the normal approximation, and/or the 
asymptotic variance of the treatment effect \cite{chow:2008, lu:2008, lu:2009,  shan:2014}.  These methods work well in large clinical trials, but generally underestimate the  size in small trials
because the normal distribution cannot adequately approximate
 the t distribution, and  the asymptotic variance underestimates the true variance of the estimated effect in ANCOVA and MMRM \cite{tang:2017}.

In this article, we propose a noniterative sample size procedure for a test based on the t distribution in finite samples.
The procedure  generalizes  Guenther's \cite{guenther:1981} method for the one sample t test and two sample t tests with equal variances, which is extended to the two sample t test
with unequal variances by Schouten \cite{schouten:1999}. 
 In Guenther's approach, the normal approximation is improved by adding  
 a correction factor. As indicated by  Schouten \cite{schouten:1999}, Guenther's  approach still underestimates the required sample size. 
We also propose a slightly more conservative sample size estimate by introducing one lower order  correction term  to Guenther's formula.
For ANCOVA and MMRM,
additional correction terms are added to account for lower order variance terms, which are functions of  covariates included in the regression.
There is limited information about the covariate distribution at the design stage  due to the inclusion/ exclusion criteria imposed on the patients.
But there is no need to specify the covariate distribution.

The proposed sample size method is suitable for superiority trials, noninferiority (NI) trials and a special case of the trials for demonstrating clinical equivalence or bioequivalence (BE). 
In Section \ref{genmeth}, we present the noniterative sample  size procedure for  a number of t tests  commonly used in the analysis of superiority trials, and assess their performance 
by simulation.
We derive accurate power formulae for ANCOVA and MMRM, and the formula for ANCOVA is exact if the covariates are  normally distributed.
Section \ref{niequiv} studies the power and sample size determination for the NI, equivalence and BE trials,
 where we also obtain the exact power for the two sample t test with unequal variance in equivalence trials. 
Numerical examples indicate that the sample size estimate (after rounding to an integer) from the noniterative procedure
 is often exact and identical to that obtained by numerically inverting the power equation.

Throughout the paper,  we let $t(f,\lambda)$ denote the t distribution with $f$ degrees of freedom (d.f.) and  noncentrality parameter $\lambda$, $t(f)$ the central t distribution,
 $F(f_1,f_2,\lambda)$ the F distribution with $f_1$ and $f_2$ d.f. and  noncentrality parameter $\lambda$, and $F(f_1,f_2)$ the central F distribution.
Let $z_p$ and $t_{f,p}$ be respectively the $p$th percentiles of the normal  $N(0,1)$ and central  $t(f)$ distributions.
Let $\Phi(\cdot)$ be the cumulative distribution function of $N(0,1)$. Let $\xv^{\otimes 2}=\xv\xv'$.

\section{A generalized sample size procedure for t tests in superiority trials}\label{genmeth}
\subsection{The generalized sample size procedure}\label{geneprocedure}
Let $\tau$ be the parameter of interest. For example, $\tau$ is the difference in the mean response between two treatment groups  in  comparative clinical trials.
Let  $\hat\tau$ be the point estimate of $\tau$,  $n^{-1}V$ the associated variance, and $\hat{V}$ the  estimate of the variance parameter $V$. Assume that
$\hat\tau$ and $\hat{V}$ are independent, and  $f \,\hat{V}/V \sim \chi_{f}^2$.  Then $\text{var}(\hat{V})=2V^2/f$ and $\text{var}(\sqrt{\hat{V}})=V/(2f)$.
Suppose we are interested in the test of equality 
\begin{equation}\label{hypoequa} 
H_0: \tau = \tau_0 \text{ \it versus } H_1: \tau = \tau_1.
\end{equation}
In comparative superiority trials, the purpose is to show that the test treatment is better than the control, and $\tau_0$ is usually set to $0$.
The test statistic $T = (\hat\tau-\tau_0)/\sqrt{n^{-1}\hat{V}} \sim t(f)$ under $H_0$. 
The null hypothesis $H_0$ is rejected if 
$|T|>t_{f,1-\alpha/2}$.

Since $T^2\sim F(1,f, n(\tau_1-\tau_0)^2/V)$  under $H_1$, 
the power of the two-sided test \eqref{hypoequa} is 
\begin{equation}\label{power00}
 P =\Pr\left[F(1, f, \frac{(\tau_1-\tau_0)^2}{n^{-1}V})> t_{f,1-\frac{\alpha}{2}}^2\right]= \Pr\left[t(f, \frac{|\tau_1-\tau_0|}{\sqrt{n^{-1}V}})> t_{f,1-\frac{\alpha}{2}}\right]  
+\Pr\left[t(f, \frac{|\tau_1-\tau_0|}{\sqrt{n^{-1}V}})<- t_{f,1-\frac{\alpha}{2}}\right],
\end{equation}
which can be well approximated by the power of the one-sided test if $\tau_1$ is not too close to $\tau_0$ to be of practical interest
\begin{equation}\label{power01}
 P\approx \Pr\left[t(f, \frac{|\tau_1-\tau_0|}{\sqrt{n^{-1}V}})> t_{f,1-\frac{\alpha}{2}}\right].
\end{equation}

The sample size is often obtained by numerically inverting Equation \eqref{power00} or by normal approximation.
The normal approximation is poor if the resulting sample size $\tilde{n}$ is small
\begin{equation}\label{sizenormal}
\tilde{n}=\frac{(z_{1-\alpha/2}+z_P)^2\,V}{(\tau_1-\tau_0)^2}.
 \end{equation}

Below we describe a generalization of Guenther's procedure  \cite{guenther:1981}  to the sample size determination for test \eqref{hypoequa}. In this approach, the sample size 
is given by
\begin{eqnarray}\label{size1}
\begin{aligned}
n_{\text{g}1}= \tilde{n} +  \frac{z_{1-\frac{\alpha}{2}}^2}{2\rho},
\end{aligned}
\end{eqnarray}
where $\rho\approx f/\tilde{n}$. If $\rho$ is a random quantity, it will be replaced  by its expected value evaluated at $\tilde{n}$. 
Guenther \cite{guenther:1981} obtained formula \eqref{size1} for the one sample t test and two sample t test with equal variance ($\rho=1$). The two sample t test with unequal variances
was studied by Schouten \cite{schouten:1999}.
 Schouten \cite{schouten:1999} indicated that formula \eqref{size1} tends to underestimate the required size for these simple t tests.
For this reason, we also propose the following slightly more conservative  estimate,
\begin{equation}\label{size2}
n_{\text{g}2}  \approx \tilde{n} +  \frac{z_{1-\frac{\alpha}{2}}^2}{2\rho} + \frac{1}{n_{\text{g}1}}  \left[\frac{z_{1-\frac{\alpha}{2}}^2}{2\rho}\right]^2=  n_{\text{g}1} + \frac{1}{n_{\text{g}1}}  \left[\frac{z_{1-\frac{\alpha}{2}}^2}{2\rho}\right]^2.
\end{equation} 
Equations \eqref{size1} and \eqref{size2} are proved in the appendix by using essentially the same argument  as  that of Schouten \cite{schouten:1999}.

We will compare formulae \eqref{size1} and \eqref{size2} with the two step (TS) procedure described in Tang \cite{tang:2017}. 
Let $f(\tilde{n})$  be the d.f. when the total  size is $\tilde{n}$. In the TS approach, the sample size is estimated as
\begin{equation}\label{sizetwo1}
 n_{_\text{TS}} = \frac{(t_{f(\tilde{n}),1-\frac{\alpha}{2}}+t_{f(\tilde{n}),P})^2 V}{(\tau_1-\tau_0)^2}.
\end{equation}

\subsection{Sample size for some commonly used t tests}
We illustrate how to use the generalized procedure in Section \ref{geneprocedure} to  calculate the power and sample size  for  the one sample t test, two sample t tests with or without equal variances, ANCOVA and MMRM.
These tests are commonly used   in the analysis of randomized clinical trials. 

\subsubsection{One sample t test}\label{onet}
Suppose $y_{i}\sim N(\mu,\sigma^2)$ for $i=1,\ldots, n$.  Let $\hat\tau=\bar{y} =n^{-1}\sum_{i=1}^n y_i$ and $\hat{V}=s^2 =(n-1)^{-1}\sum_{i=1}^n (y_i-\bar{y})^2$. 
The  test statistic can be written as 
\begin{equation*}\label{onet}
T = \frac{\hat{\tau}-\tau_0}{\sqrt{n^{-1}\hat{V}}} =\frac{\sqrt{n}(\bar{y}-\tau_0)}{\sqrt{s^2}}.
\end{equation*} 
The methods in Section \ref{genmeth} can be applied by setting $\tau_1=\mu$, $V=\sigma^2$,  $f=n-1$ and $\rho\approx 1$.  
Note that Guenther \cite{guenther:1981} obtained the noniterative sample size formula \eqref{size1}, and that formula \eqref{power00} yields the exact power for the one-sample t test.

The methods for the one sample  t test can be adapted for crossover trials without a period effect by setting 
 $\tau_1$ as the difference in two treatment means,  and $V=\sigma_d^2$, where
$y_{it}$ is the response for subject $i$ in period $t$, and $\sigma_d^2=\text{var}(y_{i1}-y_{i2})$. Please refer to Section \ref{biometh} for details.

\subsubsection{Two sample t test with equal variances}
Suppose  $y_{gi} \sim N(\mu_g,\sigma^2)$ for $i=1,\ldots, n_g$, $g=0,1$. Let $n=n_0+n_1$ be the total size, and $\gamma_g=n_g/n$ the proportion of subjects in group $g$.
Let $\bar{y}_g=n_g^{-1}\sum_{i=1}^{n_g} y_{gi}$, $s^2 = \sum_{g=0}^1\sum_{i=1}^{n_g} (y_{gi}-\bar{y}_g)^2/(n-2)$, 
$\hat\tau=\bar{y}_1-\bar{y}_0$ and  $\hat{V}=(\gamma_0^{-1}+\gamma_1^{-1}) s^2$.
The  test statistic is 
\begin{equation*}\label{onet}
T =\frac{\bar{y}_1-\bar{y}_0-\tau_0}{\sqrt{(n_0^{-1} +n_1^{-1}) s^2}} =\frac{\sqrt{n}(\bar{y}_1-\bar{y}_0-\tau_0)}{\sqrt{(\gamma_0^{-1}+\gamma_1^{-1})s^2}}.
\end{equation*} 
The methods  in Section \ref{genmeth} can be used by setting  
$\tau_1=\mu_1-\mu_0$,   $V=(\gamma_0^{-1}+\gamma_1^{-1})\sigma^2 =\frac{\sigma^2}{\gamma_0(1-\gamma_0)}$,  $f=n-2$ and $\rho\approx 1$. 
Note that Guenther \cite{guenther:1981} obtained the noniterative sample size formula \eqref{size1}, and that Equation \eqref{power00} gives the exact power.

The methods for the two sample  t test can be adapted for crossover trials with a potential period effect, where $\tau_1$ is the difference in two treatment means, 
 $\gamma_0$ is the proportion of subjects assigned to the one sequence, and
$V=\frac{\sigma_d^2}{4\gamma_0(1-\gamma_0)}$ for $\sigma_d^2$ defined in Section \ref{onet}.
 Please see Section \ref{biometh} for details.

\subsubsection{Two sample t test with unequal variances}\label{secunequalt}
Suppose $y_{gi}\sim N(\mu_g,\sigma_g^2)$. Let $\bar{y}_g=n_g^{-1}\sum_{i=1}^{n_g} y_{gi}$,  $s_g^2=\sum_{i=1}^{n_g} (y_{gi}-\bar{y}_g)^2/(n_g-1)$, 
and $\hat\tau =\bar{y}_1-\bar{y}_0$, 
$\hat{V} = s_0^2/\gamma_0 +s_1^2/\gamma_1$. The t statistic is
$$ T =\frac{\bar{y}_1-\bar{y}_0-\tau_0}{\sqrt{n_0^{-1} s_0^2+n_1^{-1} s_1^2}} =\frac{\sqrt{n}(\bar{y}_1-\bar{y}_0-\tau_0)}{\sqrt{\gamma_0^{-1} s_0^2+\gamma_1^{-1}s_1^2}}.  $$
The d.f. of the t test is computed using the Satterthwaite approximation
$$f=\frac{2\text{E}^2(n_0^{-1} s_0^2+n_1^{-1}s_1^2) }{\text{var}(n_0^{-1} s_0^2+n_1^{-1}s_1^2)} = \frac{\left(\frac{ \sigma_0^2}{n_0}+\frac{\sigma_1^2}{n_1}\right)^2}{ \frac{1}{n_0-1} \left(\frac{\sigma_0^2}{n_0}\right)^2   +\frac{1}{n_1-1} \left(\frac{\sigma_1^2}{n_1}\right)^2}.$$
The unknown $\sigma_0^2$ and $\sigma_1^2$ are  replaced respectively by $s_0^2$ and $s_1^2$ in the data analysis.

The sample size methods in Section \ref{genmeth} can be applied by setting  
 $\tau_1=\mu_1-\mu_0$,
$ V = \sigma_0^2/\gamma_0 +\sigma_1^2/\gamma_1$,  and  
$\rho=f/n\approx  V^2/(\sigma_0^4/\gamma_0^3  +\sigma_1^4/\gamma_1^3)$.  The sample size obtained by Schouten \cite{schouten:1999} is equivalent to Equation \eqref{size1}.

 Formula \eqref{power00} does not produce the exact power. The exact power can be calculated using the method of Moser {\it et al} \cite{moser:1989}.

\subsubsection{Analysis of covariance (ANCOVA)}\label{secunstratified}
Suppose in a clinical trial, $n_g$ subjects are randomized to treatment group $g$ ($g=1$ for experimental, and $0$ for placebo).  
The total sample size is $n=n_0+n_1$. 
Let $y_{gi}$  be the response, and $\xv_{gi}$ the $q\times 1 $ vector of covariates (excluding the treatment status and intercept) associated with subject $i$ in group $g$.
Let $q^*=q+2$ and $\gamma_g=n_g/n$. The data can be analyzed by the ANCOVA
\begin{equation}\label{ancova_un}
 y_{gi} \sim N(\mu + \tau g + \xv_{gi}'\betav, \sigma^2),
\end{equation}
 where $\mu$ is the intercept,  $\tau$ is the treatment effect,  $\betav$ is the covariate effect, and 
 $\sigma^2$ is the residual variance in $y_{gi}$ that is unexplained  by the covariates and treatment.

The least square estimate of the treatment effect  and its variance are given by
\begin{equation}\label{eff1}
 \hat\tau 
=\Delta_y- \Delta_x'\hat\betav \text{ and } \text{var}(\hat\tau)=\sigma^2 V_x,
\end{equation}
where $\bar{\xv}_g =n_g^{-1}\sum_{i=1}^{n_g} \xv_{gi}$, $\bar{y}_g =n_g^{-1}\sum_{i=1}^{n_g} y_{gi}$, $\Delta_{y}=\bar{y}_{1}-\bar{y}_{0}$, $\Delta_{x}=\bar{\xv}_{1}-\bar{\xv}_{0}$,  $S_{xx}=\sum_{g=0}^1 \sum_{i=1}^{n_{g}} (\xv_{gi}-\bar{\xv}_{g})^{\otimes 2}$,
 $S_{xy}=\sum_{g=0}^1 \sum_{i=1}^{n_{g}} (\xv_{gi}-\bar{\xv}_{g})y_{gi}$, 
$\hat\betav = S_{xx}^{-1} S_{xy}$,  $\Upsilon =n \gamma_0 \gamma_1 \Delta_x'S_{xx}^{-1}\Delta_x$ and 
$ V_x=n_0^{-1}+n_1^{-1}+\Delta_x' S_{xx}^{-1}\Delta_x = (1+ \Upsilon)/(n\gamma_0\gamma_1)$.
Let $f=n-q^*$ and $\hat\sigma^2=f^{-1}\sum_{g=0}^1\sum_{j=1}^{n_g}[y_{gj}- \bar{y}_g-(\xv_{gj}-\bar{\xv}_g)'\hat\betav]^2$.
In ANCOVA, the inference is made by assuming $\xv_{gi}$'s  are known and fixed. Given  $\xv_{gi}$'s, the test statistic for $H_0:\tau=\tau_0$ is distributed as
\begin{equation*}\label{testancova_un}
 T = \frac{\hat\tau-\tau_0}{ \sqrt{\hat\sigma^2 V_x}} \sim t\left[f, \frac{\tau_1-\tau_0}{\sqrt{\sigma^2 V_x}}\right] \text{ and } T^2\sim F\left[1,f, \frac{(\tau_1-\tau_0)^2}{\sigma^2 V_x}\right].
\end{equation*} 

At the design stage, $\xv_{gi}$'s are typically unknown.   The power   is given by
\begin{eqnarray}\label{power_ancova_un}
\begin{aligned}
P =\int \text{Pr}\left[ F\left(1,f,\frac{(\tau_1-\tau_0)^2}{\sigma^2V_x(\tilde{\Upsilon})}\right) > t_{f,1-\frac{\alpha}{2}}^2 \right] g(\tilde{\Upsilon})d \tilde{\Upsilon}
\approx \int \text{Pr}\left[ t\left(f,\sqrt{\frac{(\tau_1-\tau_0)^2} {\sigma^2V_x(\tilde{\Upsilon})}}\right) > t_{f,1-\frac{\alpha}{2}} \right] g(\tilde{\Upsilon})d \tilde{\Upsilon},
\end{aligned}
\end{eqnarray}
where  $g(\tilde{\Upsilon})$ is the probability density function (PDF) of $\tilde{\Upsilon}= (n-1-q)\Upsilon/q$, and $V_x(\tilde{\Upsilon})=\frac{ 1+ q \tilde{\Upsilon}/(n-q -1)}{n\gamma_0\gamma_1}$.
We assume $\tilde{\Upsilon} \sim F(q,n-q-1)$. The assumption holds exactly, and Equation \eqref{power_ancova_un} yields the exact power 
if $\xv_{gi}$ is normally distributed \cite{tang:2017}. 
 For nonnormal covariates, the power estimation based on the approximation $\tilde{\Upsilon} \sim F(q,n-q-1)$ 
generally leads to very accurate power estimate  in randomized trials (i.e.  no systematic difference in the  distribution of  $\xv_{gi}$ between two groups), and 
this  will be demonstrated in Section $4$.
To avoid numerical integration, we  approximate Equation \eqref{power_ancova_un} by replacing $\tilde{\Upsilon}$ by $\text{E}(\tilde{\Upsilon}) \approx (n-1-q)/(n-3-q)$ 
\begin{eqnarray}\label{ancovapower_unapp1}
\begin{aligned}
 P &\approx\text{Pr}\left[ F\left(1,f,\frac{ n\gamma_0\gamma_1(\tau_1-\tau_0)^2}{\sigma^2 (1+ \frac{q}{n-q-3})}\right) > t_{f,1-\frac{\alpha}{2}}^2 \right] \approx \text{Pr}\left[ t\left(f,\sqrt{\frac{ n\gamma_0\gamma_1(\tau_1-\tau_0)^2}{\sigma^2 (1+ \frac{q}{n-q-3})}}\,\right) > t_{f,1-\frac{\alpha}{2}} \right].
\end{aligned}
\end{eqnarray}

In large trials, the sample size is commonly estimated based on the normal approximation and the asymptotic variance $\text{var}(\hat\tau)\approx\sigma^2/(n\gamma_0\gamma_1)$
\begin{equation}\label{normancova_asy}
   n_{\text{asy}} = \frac{ (z_{1-\frac{\alpha}{2}}+z_{P})^2 \sigma^2 }{\gamma_0\gamma_1(\tau_1-\tau_0)^2}.
\end{equation}
Another common approach is to invert the  power formula below based on the t distribution and  asymptotic variance \cite{shan:2014},
\begin{eqnarray}\label{ancovapower_unapp_t}
\begin{aligned}
 P &\approx\text{Pr}\left[ F\left(1,f,\frac{ n\gamma_0\gamma_1(\tau_1-\tau_0)^2}{\sigma^2}\right) > t_{f,1-\frac{\alpha}{2}}^2 \right] \approx \text{Pr}\left[ t\left(f,\sqrt{\frac{ n\gamma_0\gamma_1(\tau_1-\tau_0)^2}{\sigma^2}}\,\right) > t_{f,1-\frac{\alpha}{2}} \right],
\end{aligned}
\end{eqnarray}
and it yields slightly better performance than  Borm {\it et al} \cite{borm:2007} approach, in which the total sample size from the normal approximation  \eqref{normancova_asy} is inflated by $2$ (i.e. $1$ subject per arm).

The sample size based on the normal approximation and the exact variance is
\begin{equation}\label{sizelowunstra}
\tilde{n}= \frac{ (z_{1-\frac{\alpha}{2}}+z_{P})^2 \sigma^2 \,\text{E}(V_x) }{(\tau_1-\tau_0)^2}= n_{\text{asy}}\left[1+ \frac{q}{\tilde{n}-q-3}\right],
\end{equation}
The solution to Equation \eqref{sizelowunstra} is given in the  appendix, and it satisfies $ n_{\text{asy}} + q<\tilde{n} <n_{\text{asy}} + q+3$.
Inserting $\tilde{n}\approx n_{\text{asy}} +q+1$ into the last term in Equation \eqref{sizelowunstra} gives
\begin{equation}\label{normancova1}
\tilde{n}\approx   n_{\text{asy}} \left[1+\frac{q}{ n_{\text{asy}} -2}\right].
\end{equation}
Plugging $\tilde{n}$ into Equations \eqref{size1} and \eqref{size2} yields the size based on the t distribution ($\rho=1$). 
We use the approximation \eqref{normancova1} instead of the explicit solution to Equation \eqref{sizelowunstra} to slightly simplify the  calculation. It also
 enables the generalization of the method to MMRM that will be investigated in Section \ref{secmmrm}.

In the two step approach,   Equation \eqref{sizetwo1}  is calculated as
\begin{equation*}\label{sizeupunstra3}
 n_u \approx n_{u_\text{asy}} \left[1+\frac{q}{ n_{u_\text{asy}} -2}\right],
\end{equation*}
where $n_{u_\text{asy}} = (t_{\tilde{n}-q^*,1-\alpha/2}+t_{\tilde{n}-q^*,P})^2 \sigma^2 /[\gamma_0\gamma_1(\tau_1-\tau_0)^2]$.

\subsubsection{Mixed effects model for repeated measures (MMRM)}\label{secmmrm}
Suppose in a  clinical trial,  $n$ subjects are randomly assigned to the experimental ($g=1$) or control ($g=0$) treatment.
Let $n_g$ and $\gamma_g=n_g/n$ be the number and proportion of subjects randomized to  group $g$. 
Let $\yv_{gi}=(y_{gi1},\ldots,y_{gip})'$ be
the outcomes  collected at $p$ post-baseline visits, and $\xv_{gi}$ the $q\times 1$ vector of  covariates for subject $i$ in group $g$.
Let $q^*=q+2$.
In clinical trials, the data are missing mainly due to dropout \cite{tang:2017}.
At the design stage, it is reasonable to assume the  missing data pattern is monotone in the sense that
  if $y_{gij}$ is observed, then $y_{git}$'s are observed for all $t\leq j$.
Let $n_{gj}$ and $\pi_{gj}= n_{gj}/n_g$ be the number and proportion of subjects  retained at visit $j$  in group $g$.
The total number of subjects retained at visit $j$ is $m_j=\sum_{g=0}^1 n_{gj}$, and
the pooled retention rate at  visit $j$ is $\bar\pi_j=\sum_{g=0}^1 \gamma_g\pi_{gj}$.
Without loss of generality, we sort the data so that within each group, 
subjects who stay in the trial longer will have smaller index $i$ than subjects who discontinue earlier.

The following MMRM is often used to analyze longitudinal  clinical data collected at a fixed number of timepoints \cite{aiddiqui:2009, laird:1987}
\begin{equation}\label{mixed1}
\yv_{gi} \sim N_p[(\mu_1+\alphav_1' \xv_{gi}+\tau_1g,\ldots,\mu_p+\alphav_p' \xv_{gi}+\tau_pg)',\Sigma].
\end{equation} 
where $\Sigma$ is an unstructured (UN) covariance matrix. A structured covariance matrix (possibly induced via the use of random effects)
can be useful when individuals have   a large number of observations, or varying time points of observations \cite{laird:1987}.
In MMRM, inference is often made based on the restricted maximum likelihood (REML) and Kenward-Roger \cite{1997:kenward} adjusted
 variance estimate to reduce the small sample bias \cite{tang:2017}.

Let $\Sigma=L\Lambda L'$ be the LDL decomposition of $\Sigma$, where $U={\small \begin{bmatrix} 1 & 0 &  \ldots & 0\\
                       -\beta_{21} & 1 & \ldots &0 \\
                                       & \ldots & \ldots & 0\\
                      -\beta_{p1} & \ldots & -\beta_{p,p-1} & 1 \\
                    \end{bmatrix}}$,   $L=U^{-1}$ and $\Lambda=\text{diag}(\sigma_1^2,\ldots,\sigma_p^2)$.  Let $l_{jt}$ be the $(j,t)$-th entry of $L$.  
Model \eqref{mixed1} can be reorganized as the product of the following simple regression models \cite{2016:tang,tang:2017e}
\begin{equation}\label{factor} 
y_{gij}= \zv_{gij}'\thetav_j + \varepsilon_{gij} \text{ for } j\leq p,
\end{equation}
 where  $(\underline\mu_j,\underline\alphav_{j}',\underline\tau_j)'=(\mu_j,\alphav_{j}',\tau_j)' - \sum_{t=1}^{j-1} \beta_{jt} (\mu_t,\alphav_{t}',\tau_t)' $, 
$\betav_j=(\beta_{j1},\ldots,\beta_{j,j-1})'$, 
 $\thetav_j=(\underline\mu_j,\underline\alphav_{j}',\underline\tau_j, \betav_j')'$, 
    $\zv_{gij} =(1,\xv_{gi}', g,y_{gi1},\ldots,y_{gi,j-1})'$, and  $\varepsilon_{gij} \stackrel{i.i.d.}{\sim} N(0,\sigma_j^2)$. 

Tang \cite{tang:2017} derives the REML estimate for model \eqref{factor}, and studies its theoretical properties
\begin{equation*}\label{lsest}
\hat\thetav_j =( \sum_{g=0}^1 \sum_{i=1}^{n_{gj}}\zv_{gij}\zv_{gij}' )^{-1}\sum_{g=0}^1 \sum_{i=1}^{n_{gj}} \zv_{gij}y_{gij}  \text{ and }
 \hat\sigma_j^2=\frac{\sum_{g=0}^1 \sum_{i=1}^{n_{gj}} (y_{gij}- \zv_{gij}'\hat\thetav_j)^2}{m_j-q^*}.
\end{equation*}
The treatment effect estimate at visit $p$ is $\hat\tau_p=\sum_{j=1}^p \hat{l}_{pj} \hat{\underline\tau}_j$,
 and its  Kenword-Roger variance  estimate  is
\begin{equation}\label{varkr}
\widehat{\textrm{var}}(\hat\tau_p) = \sum_{j=1}^p \hat{l}_{pj}^2\hat\sigma_j^2 V_{x_j} +
           2 \sum_{j=2}^{p} \hat{l}_{pj}^2 \hat\sigma_j^2 \frac{\sum_{t=1}^{j-1} [V_{x_j}-V_{x_t}]}{m_j-q^*},
\end{equation}
where  $\bar\xv_{{gj}}=n_{gj}^{-1}\sum_{i=1}^{n_{gj}} \xv_{gi}$,
 $\Delta_{j}=\bar{\xv}_{1j}-\bar{\xv}_{0j}$, $S_{x_j}=\sum_{g=0}^1 \sum_{i=1}^{n_{gj}} (\xv_{gi}-\bar{\xv}_{gj})^{\otimes 2}$, and  
 $V_{x_j}= n_{1j}^{-1}+n_{0j}^{-1}+\Delta_{j}' S_{x_{j}}^{-1}\Delta_{j}$.
 
We use slightly different notation in MMRM. We let $\tau_1$ denote the treatment effect at first timepoint. The true value
for $\tau_j$ under $H_1$ is $\tau_{j1}$, and its value under $H_0$ is $\tau_{j0}$. 
The test statistic for $H_0: \tau_p=\tau_{p_0}$ {\it vs} $H_1:\tau_p=\tau_{p1}$, 
$$ T =\frac{\hat\tau_p-\tau_{p_0}}{ \sqrt{\widehat{\textrm{var}}(\hat\tau_p)  }} $$
approximately follows a  $t$  distribution under $H_0$, and the d.f. is obtained from the Satterthwaite approximation \cite{1997:kenward}
\begin{eqnarray}\label{mmrmdf}
\begin{aligned}
\hat{f}=  \frac{2\widehat{\text{E}}^2( \sum_{j=1}^p \hat{l}_{pj}^2\hat\sigma_j^2 V_{x_j})}{\widehat{\text{var}}( \sum_{j=1}^p \hat{l}_{pj}^2\hat\sigma_j^2 V_{x_j})} =\frac{ (\sum_{j=1}^p \hat{l}_{pj}^2 \hat\sigma_j^2 V_{x_j})^2}{2\sum_{j=2}^pA_j + \sum_{j=1}^p \frac{ \hat{l}_{pj}^2 a_j^2 }{m_j-q^*} },
\end{aligned}
\end{eqnarray}
where $a_j=\hat{ l}_{pj}\hat\sigma_j^2 V_{x_j}$, $\av_j=(a_1,\ldots,a_{j-1})'$ and $A_j= \hat{l}_{pj}^2 \hat{\sigma}_j^2  \av_j'\hat{L}_{j-1}' (Y_j'Q_jY_j)^{-1} \hat{L}_{j-1}\av_j$,
$Y_{j}$ and $X_{j}$ are $m_j\times (j-1)$ and $m_j\times q^*$ matrices
whose $(n_{0j}g+i)$-th rows contain $(y_{gi1},\ldots,y_{gi,j-1})$ and $(1,\xv_{gi}',g)$ respectively, and $Q_j=I-X_{j}(X_{j}'X_{j})^{-1}X_{j}$.
The derivation of Equation \eqref{mmrmdf} and two other equations (\eqref{vartau} and \eqref{normfrho}) below is given in the appendix. 

Lu {\it et al} \cite{lu:2008, lu:2009} developed power and sample size methods for MMRM. These methods are based on the asymptotic variance of $\hat\tau_p$ instead of the commonly used 
Kenword-Roger adjusted variance estimate. The Kenword-Roger variance  estimate \cite{1997:kenward} provides a roughly unbiased estimate of the variance $V_{\tau}$ of $\hat\tau_p$ while ignoring the lower order term
\begin{equation}\label{vartau}
V_{\tau}=  \sum_{j=1}^p l_{pj}^2\sigma_j^2  V_{x_j} + \sum_{j=2}^p l_{pj}^2  \sum_{t=1}^{j-1} \omega_{jt} \sigma_t^2 [V_{x_j}-V_{x_t}].
\end{equation}
where  $\omega_{jt} = \sigma_j^2/[(m_j-q^*-j)\sigma_t^2]$ 
is the $(t,t)$-th element of $L_{j-1}' \text{\normalfont var}(\hat\betav_j)  L_{j-1}$.

In the MMRM analysis, $\xv_{gi}$'s are assumed to be fixed, but unknown at the design stage.  In the power calculation, we will replace  $V_{x_j}$'s, $\hat{f}$ and $\widehat{\textrm{var}}(\hat\tau_p)$  by their expected values
$$\tilde{V}_{x_j}  = \text{E}[V_{x_j}]= \frac{\varpi_{\pi_t}}{n} [1 + \frac{q}{n\bar\pi_t-q-3}],$$
$$V_{\tau}^*=\text{E}[\widehat{\textrm{var}}(\hat\tau_p)]=  \sum_{j=1}^p c_j \tilde{V}_{x_j} +
           2 \sum_{j=2}^{p} c_j \frac{\sum_{t=1}^{j-1} (\tilde{V}_{x_j}-\tilde{V}_{x_t})}{m_j-q^*},$$
\begin{equation}\label{normfrho}
f=\text{E}(\hat{f}) \approx \frac{ (\sum_{j=1}^p c_j \tilde{V}_{x_j})^2}{2\sum_{j=2}^p c_j \frac{ \sum_{t=1}^{j-1} c_t \tilde{V}_{x_t}^2}{m_j-q^*-j }+ \sum_{j=1}^p \frac{ c_j^2\tilde{V}_{x_j}^2 }{m_j-q^*} },
\end{equation}
where $\varpi_{\pi_t}= \sum_{g=0}^1 (\gamma_g\pi_{gt})^{-1}$ and 
$c_j= \text{E}(\hat{ l}_{pj}^2\hat\sigma_j^2) = (1-\frac{j-1}{m_j-q^*})[l_{pj}^2\sigma_j^2 +\sum_{k=j+1}^p \frac{1}{m_k-q^*-k} l_{pk}^2\sigma_k^2]$.
It is possible to  derive a better approximation of the d.f. $\text{E}(\hat{f})$. We will not pursue it further here.

The power of the Wald test at  a two-sided  significance level of $\alpha$ is given by 
\begin{eqnarray}\label{power0mmrm}
\begin{aligned}
P= \text{Pr}\left[F\left(1,f, \frac{(\tau_{p_1}-\tau_{p_0})^2}{ V_{\tau}^*}\right) > t_{f,1-\frac{\alpha}{2}}^2 \right]  \approx \text{Pr}\left[t\left(f,\frac{|\tau_{p_1}-\tau_{p_0}|}{ \sqrt{V_{\tau}^*}}\right) \geq  t_{f,1-\frac{\alpha}{2}}\right].
\end{aligned}
\end{eqnarray}
One may approximate $V_{\tau}^*$ by $V_{\tau}$, and/or  $f$ by  $f_o=(m_1-q^*)\rho_o$ to simplify the calculation, where
$\rho_o=  \sum_{j=1}^p l_{pj}^2\sigma_j^2 \tilde{V}_{x_1} /\sum_{j=1}^p l_{pj}^2\sigma_j^2  \tilde{V}_{x_j} $ 
can be  interpreted as the fraction of observed information among subjects retained at visit $1$.
The following approximation of Tang \cite{tang:2017} is only slightly less accurate than Equation \eqref{power0mmrm}  even in small samples
\begin{eqnarray}\label{power1mmrm}
\begin{aligned}
P= \text{Pr}\left[F\left(1,f_o, \frac{(\tau_{p_1}-\tau_{p_0})^2}{V_{\tau}}\right) > t_{f_o,1-\frac{\alpha}{2}}^2 \right] \approx \text{Pr}\left[t\left(f,\frac{|\tau_{p_1}-\tau_{p_0}|}{ \sqrt{V_{\tau}}}\right) \geq  t_{f_o,1-\frac{\alpha}{2}}\right].
\end{aligned}
\end{eqnarray}

The sample size based on the normal approximation and the asymptotic variance is given by 
\begin{equation}\label{size0mmrm}
n_{\text{a}}= \frac{ (z_{1-\frac{\alpha}{2}}+z_{P})^2 \sum_{j=1}^p l_{pj}^2 \sigma_j^2\varpi_{\pi_j}}{(\tau_{p_1}-\tau_{p_0})^2}.
\end{equation} 
The sample size based on the normal approximation and the variance defined in Equation \eqref{vartau} is given by
\begin{equation}\label{size1mmrm}
\tilde{n}= \frac{ (z_{1-\frac{\alpha}{2}}+z_{P})^2 V_{\tau}}{(\tau_{p_1}-\tau_{p_0})^2} \approx n_{\text{a}} \sum_{j=1}^p b_j \left[d_j  +
       \frac{e_j}{n_{\text{a}}\bar\pi_j-j+1} \right].
\end{equation} 
where $b_j= l_{pj}^2 \sigma_j^2\varpi_{\pi_j}/\sum_{j=1}^p l_{pj}^2 \sigma_j^2\varpi_{\pi_j}$, 
$d_j=1+q/(n_{\text{a}}\bar\pi_j-2)$, and $e_j=\sum_{t=1}^j (d_j-\varpi_{\pi_t}d_t /\varpi_{\pi_j})$.
To derive \eqref{size1mmrm}, we assume  $n\bar\pi_j-q-1 \approx n_{\text{a}}\bar\pi_j$  
by the same argument  as that for Equation \eqref{normancova1}.

Plugging $\tilde{n}$ and $\rho=f/(\tilde{n}\,\bar{\pi}_1 -q^*)$ into Equations \eqref{size1} and \eqref{size2} yields the size based on the t distribution, where
$f$ is estimated using Equation \eqref{normfrho} at $n=\tilde{n}$.

In the TS procedure, the sample size is  calculated as
\begin{equation*}\label{sizeupunstra3}
 n_u 
\approx n_{u_\text{a}} \sum_{j=1}^p b_j \left[d_j^*  +
       \frac{\sum_{t=1}^{j}(d_j^* - \varpi_{\pi_t}d_t^*/\varpi_{\pi_j})}{n_{u_\text{a}}\bar\pi_j-(j-1)} \right].
\end{equation*}
where $n_{u_\text{a}} = (t_{f_l,1-\alpha/2}+t_{f_l,P})^2  \sum_{j=1}^p l_{pj}^2 \sigma_j^2\varpi_{\pi_j} /(\tau_{p_1}-\tau_{p_0})^2$, $d_j^*=1+q/(n_{u_\text{a}}\bar\pi_j-2)$, and
$f_l=(\tilde{n}-q^*)\rho$. It slightly improves the TS procedure described in \cite{tang:2017}.

\subsection{Numerical Examples}
We present three numerical examples to assess the performance of the proposed methods  in superiority trials. 

\begin{example}\label{ttestexam}
\normalfont
Table \ref{ttestres} displays the sample size estimates for the two sample t test using the exact method, normal approximation, TS approach and the noniterative 
 method. The variance is set to $\sigma^2=1$ for the test assuming equal variances, and $(\sigma_0^2, \sigma_1^2)=(1,4)$ in the  test with unequal variances. 
Other parameters are set as $\gamma_0=\gamma_1=1/2$, and  $\tau_1=\mu_1-\mu_0=0.5,0.75,1,1.25,1.5,1.75, 2, 2.25$. 

The sample size estimate is not rounded to an integer value for the purpose of comparison. 
The normal approximation underestimates the sample size in all cases. The TS procedure produces slightly conservative size estimates particularly at large $\mu_1-\mu_0$.
Although Equation \eqref{size1} is more accurate than the normal approximation, it still underestimates the sample size. 
The sample size estimate from Equation \eqref{size2} is surprisingly close to the exact value in all cases.

In practice, the sample size must take an integer value. 
Equation \eqref{size2} yields the same  estimate (after rounding  to integers) as the exact method in all cases.
 Equation \eqref{size1} underestimates the required size  at $\mu_1-\mu_0=2.0$ (simulated power based on $1,000,000$ trials is $79.05\%$ at $n=10$; exact power is $79.05\%$) and $1.5$
(simulated power is $79.68\%$ at $n=16$; exact power is $79.65\%$) for the tests with equal variances
 even after the sample size estimate is rounded up to the next integer.

We simulate $1,000,000$ trials.  The exact sample size per treatment arm is  rounded up to the nearest integer. 
The simulated power (SIM) is close to the nominal power in all cases. This is expected since
there is more than $95\%$ chance that the simulated power (standard error $\leq 0.04\%$) lies within $0.08\%$ of the true power. 
\end{example}

 \begin{table}[h]
\begin{center}
\caption{Sample size needed to achieve $80\%$ power at the two-sided significance level of $\alpha=0.05$ for the two sample t tests:\newline
$^{(a)}$ The sample size estimate is not rounded  to the nearest integer for the purpose of comparison;\newline
$^{(b)}$ The exact power and sample size are calculated by using Equation \eqref{power01} for  tests with equal variances, and by Moser {\it et al} \cite{moser:1989} method for the tests with unequal variances;\newline
$^{(c)}$  The exact size per arm is rounded up to the nearest integer;\newline
$^{(d)}$  Simulated power (SIM) based on $1,000,000$ simulated trials. 
}\label{ttestres}
\begin{tabular}{ccrrrc@{\extracolsep{5pt}}c@{}c@{}ccccccc} \\\hline 
    &        \multicolumn{5}{c}{estimated total  size$^{(a)}$ at $\alpha=0.05, P=80\%$} &  \\ \cline{2-6}
                        &                        &&  two & \multicolumn{2}{c}{noniterative} &    \multicolumn{3}{c}{power ($\%$)} \\\cline{5-6}\cline{7-9} 
$\mu_1-\mu_0$  &  exact$^{(b)}$  & normal & step & \eqref{size1} &  \eqref{size2} &  size$^{(c)}$  & exact$^{(b)}$ &   SIM$^{(d)}$ \\\hline
\multicolumn{6}{c}{ two sample t test with equal variances}\\
                                                             $0.50$&$127.53$&$125.58$&$127.59$&$127.50$&$127.53$ & $64$ & $80.15$ & $80.18$ \\
                                                                $0.75$&$ 57.80$&$ 55.81$&$ 57.90$&$ 57.73$&$ 57.80$ & $29$ & $80.14$ & $80.16$ \\
                                                                  $1.00$&$ 33.43$&$ 31.40$&$ 33.59$&$ 33.32$&$ 33.43$ & $17$ & $80.70$ & $80.75$\\
                                                                  $1.25$&$ 22.19$&$ 20.09$&$ 22.46$&$ 22.01$&$ 22.18$ & $12$ & $83.30$ & $83.30$\\
                                                                  $1.50$&$ 16.12$&$ 13.95$&$ 16.56$&$ 15.87$&$ 16.11$ & $9$ & $84.76$ & $84.76$\\
                                                                  $1.75$&$ 12.50$&$ 10.25$&$ 13.22$&$ 12.17$&$ 12.48$ & $7$ & $85.16$  & $85.12$\\
                                                                  $2.00$& $ 10.18$&$  7.85$&$ 11.36$&$  9.77$&$ 10.15$ & $6$ & $87.64$ & $87.64$\\
                                                               $2.25$&$  8.62$&$  6.20$&$ 10.59$&$  8.12$&$  8.58$ & $5$ & $87.46$ & $87.49$ \\

\vspace*{2pt}\\
\multicolumn{6}{c}{ two sample t test with unequal variances}\\
$0.50$&$316.59$&$313.96$&$316.64$&$316.57$&$316.59$ & $159$ & $80.18$ & $80.13$\\                                                                                                                                     
$0.75$&$142.19$&$139.54$&$142.27$&$142.15$&$142.20$ & $72$ & $80.50$ & $80.52$\\                                                                                                                                     
$1.00$&$ 81.18$&$ 78.49$&$ 81.29$&$ 81.10$&$ 81.19$ & $41$ & $80.40$ & $80.42$\\                                                                                                                                     
$1.25$&$ 52.97$&$ 50.23$&$ 53.12$&$ 52.85$&$ 52.97$ & $27$ & $80.79$ & $80.84$\\                                                                                                                                     
$1.50$&$ 37.68$&$ 34.88$&$ 37.89$&$ 37.50$&$ 37.68$ & $19$ & $80.36$ & $80.36$\\                                                                                                                                     
$1.75$&$ 28.49$&$ 25.63$&$ 28.79$&$ 28.24$&$ 28.48$ & $15$ & $82.21$ & $82.17$ \\                                                                                                                                     
$2.00$&$ 22.55$&$ 19.62$&$ 22.97$&$ 22.23$&$ 22.54$ & $12$ & $82.74$ & $82.72$\\                                                                                                                                     
$2.25$&$ 18.51$&$ 15.50$&$ 19.10$&$ 18.12$&$ 18.49$ & $10$ &$83.52$ & $ 83.47$\\
\hline
 \end{tabular} 
\end{center}
\end{table}

\begin{example}\label{ancovaexample}
\normalfont
We assess the power and sample size  formulae for ANCOVA based on  two models. 
In Model $1$,  the baseline outcome $x_{gi} \sim N(0,1)$ is used as the covariate  ($q=1$), and  $y_{gi} \sim N(0.5+\tau g +0.5 x_{gi},\sigma^2)$, where
 $\gamma_0=\gamma_1=1/2$, $\sigma^2=1$ and $\tau=1,1.25, 1.5, 1.75, 2$.

In Model $2$, the covariates ($q=3$) include
 the baseline outcome $x_{gi}$, and a categorical prognostic factor $A$ with three levels, and
$y_{gi} \sim N(\eta_s+\tau g +0.5 x_{gi},\sigma^2)$ for subjects in level $s$ of $A$, where $\eta_1=0.5, \eta_2=0, \eta_3=1$.
Subjects are in level $1$, $2$, and $3$ of factor $A$ with probability $0.4$, $0.4$, and $0.2$ respectively. Other setup is the same as Model $1$.
The power and sample size depend on $(\tau, \sigma^2, \gamma_0,\gamma_1, q)$. Other parameters are specified for data simulation.

 Table \ref{ancovares} reports the results. The sample size per arm is calculated by numerical inversion of Formula \eqref{power_ancova_un} at the  $80\%$ power, and rounded up to the nearest integer.
We simulate $1,000,000$ trials. 
For Model 1 with normally distributed  covariates, Formula \eqref{power_ancova_un} yields the exact power estimate.
It also produces very accurate power estimate  for Model 2 with nonnormal covariates, which 
are within $0.1\%$ of the simulated power in all cases. 
Formula \eqref{ancovapower_unapp1} is generally accurate. Its performance slightly deteriorates when the number of covariates  is relatively large in small samples. 
In the worst case ($q=3$, $n_0=n_1=7$ subjects per arm), the estimate by Equation \eqref{ancovapower_unapp1} deviates from the simulated power by $0.59\%$.

We compare several sample size methods.  The normal approximation can substantially underestimate the sample size. 
 For example, when $\tau=2$ in Model 2,
the target size is $13.66$ while Equations \eqref{normancova_asy} and \eqref{normancova1} yield the estimates of $7.85$ and   $11.87$ respectively. 
As a rule of thumb, the sample size will be underestimated by  about  $q+  z_{1-\alpha/2}^2/2$ by Equation \eqref{normancova_asy}, and by about $z_{1-\alpha/2}^2/2$ by Equation \eqref{normancova1}. 
The method by  inverting Equation \eqref{ancovapower_unapp_t} underestimates the size by about $q$ in all cases.
Formulae \eqref{size1} and  \eqref{size2} and the TS procedure generally yield accurate  size estimates.
The estimate from Equation \eqref{size2} tends to be the closest to the exact size except when the number of covariates is large and 
the total sample size is small.
\end{example}

 \begin{table}[htbp]
\begin{center} \caption{Calculated sample sizes and   power estimates  for ANCOVA: \newline
$^{(a)}$ The estimate  is exact for normally distributed covariates (i.e. $q=1$). The  per arm size is rounded up to the nearest integer;\newline
$^{(b)}$ Simulated power (SIM) based on $1,000,000$ simulated trials. }\label{ancovares}
\begin{tabular}{ccccccccccccccccccccccccccccc} \\\hline 
    &       \multicolumn{7}{c}{estimated total size at $P=80\%, \alpha=0.05$ } && size & \multicolumn{3}{c}{power ($\%$) }\\ \cline{2-8}\cline{11-13}
           &    inversion      & \multicolumn{2}{c}{normal} & inversion  & two  & \multicolumn{2}{c}{noniterative}    && per &  & \multicolumn{2}{c}{nominal}         \\ \cline{3-4} \cline{7-8} \cline{12-13}
$\tau$  & \eqref{power_ancova_un}$^{(a)}$ & \eqref{normancova_asy} & \eqref{normancova1} & \eqref{ancovapower_unapp_t}& step & \eqref{size1} &  \eqref{size2}  &&   arm $^{(a)}$ &  SIM$^{(b)}$ & \eqref{power_ancova_un} & \eqref{ancovapower_unapp1}   \\\hline
\multicolumn{12}{c}{ $q=1: y_{gj} \sim N(\mu+\tau g +0.5 x_{gj},1)$}\\
                                  $1.00$&$ 34.50$&$ 31.40$&$ 32.46$& $33.50$  &$ 34.65$&$ 34.38$&$ 34.49$&&$ 18$&$81.85$&$81.80$&$81.79$\\
                                 $1.25$&$ 23.30$&$ 20.09$&$ 21.20$&  $22.30$ &  $ 23.54$&$ 23.12$&$ 23.28$&&$ 12$&$81.34$&$81.34$&$81.30$\\
                              $1.50$&$ 17.26$&$ 13.95$&$ 15.12$&  $16.28$ & $ 17.66$&$ 17.04$&$ 17.26$&&$  9$&$81.96$&$82.00$&$81.93$\\
                       $1.75$&$ 13.67$&$ 10.25$&$ 11.49$& $12.72$ & $ 14.30$  & $ 13.41$&$ 13.69$&&$  7$&$81.20$&$81.25$&$81.07$\\
                  $2.00$&$ 11.37$&$  7.85$&$  9.19$& $10.47$ & $ 12.32$&$ 11.11$&$ 11.44$&&$  6$&$82.92$&$82.96$&$82.74$\\

\vspace*{2pt}\\
\multicolumn{12}{c}{ $q=3:  y_{gj} \sim N(\eta_s+\tau g +0.5 x_{gj},1)$ category $s=1,2,3$}\\
                  $1.00$&$ 36.64$&$ 31.40$&$ 34.60$&$33.64$  &$ 36.77$&$ 36.52$&$ 36.62$&&$ 19$&$81.72$&$81.64$&$81.61$\\
                $1.25$&$ 25.49$&$ 20.09$&$ 23.42$&$22.54$ & $ 25.71$&$ 25.35$&$ 25.49$&&$ 13$&$80.99$&$80.98$&$80.88$\\
              $1.50$&$ 19.49$&$ 13.95$&$ 17.46$& $16.66$ &$ 19.86$&$ 19.38$&$ 19.57$&&$ 10$&$81.40$&$81.38$&$81.18$\\
            $1.75$&$ 15.93$&$ 10.25$&$ 13.98$&$13.26$ & $ 16.48$&$ 15.90$&$ 16.13$&&$  8$&$80.18$&$80.25$&$79.78$\\
             $2.00$&$ 13.66$&$  7.85$&$ 11.87$& $11.19$ & $ 14.39$&$ 13.80$&$ 14.06$&&$  7$&$81.59$&$81.61$&$81.00$\\
\hline
 \end{tabular}
\end{center}
\end{table}

\begin{example}\label{mmrmexample}
\normalfont
We revisit the sample size estimation based on MMRM in the design of a new antidepressant trial investigated by Tang \cite{tang:2017}. The primary objective of the trial is to assess the effect of a new compound 
on depression. The Hamilton $17$-item rating scale for depression ($\text{HAMD}_{17}$)  will be  collected at baseline 
and $p=4$ post-randomization visits. Suppose
$$\begin{bmatrix} y_{gi1} \\
 y_{gi2}\\
y_{gi3}\\
y_{gi4}\end{bmatrix} \sim N\left( \begin{bmatrix} \mu_{gi1} \\ \mu_{gi2} \\ \mu_{gi3} \\ \mu_{gi4} \\\end{bmatrix}, 
 \begin{bmatrix} 19.68  & 16.45  &  15.39  &   16.36 \\
             16.45    &    34 &     25.34 &    26.13 \\
             15.39    & 25.34   &  38.44 &    33.91\\
            16.36    &  26.13 &     33.91  &   45.28  \\ \end{bmatrix}\right),$$
and the retention rate is $(\pi_{01},\ldots,\pi_{04})= (1,0.92,0.86,0.74)$ and $(\pi_{11},\ldots,\pi_{14})=(1,0.93,0.87,0.76)$,
where $\mu_{gi1}=3.3+0.72y_{gi0} +0.1g$, $\mu_{gi2}=2.7+0.69y_{gi0}-1.5g$,
$\mu_{gi3}=2.9+0.61y_{gi0}-2.3g$, $\mu_{gi4}=1+0.67y_{gi0}+\tau_4 \,g$.
These parameters are specified based roughly on the MMRM analysis of an antidepressant trial.
The  sample size
depends on $(\tau_4,\Sigma, \pi_{gj}'s, q)$. Other parameters are specified in order to simulate the data.

We set $\tau_4=-4$, $-8$ or $-12$. Three alternative covariance structures are considered to serve  as the sensitivity analysis:
1)   a compound symmetry (CS) structure: $\Sigma_{kk}=45$ and $\Sigma_{kj}=15$ if $k\neq j$; 2) 
 an autoregressive (AR(1)) structure $\Sigma_{jk}= 45\times 0.8^{|j-k|}$; 3)  a Toeplitz (TO)  structure $\Sigma_{jk}=40 -6|j-k|$.

We calculate the total  size needed to achieve $90\%$ power at  $\alpha=0.05$ using Equation \eqref{size2}, which is rounded up to the nearest integer.
The size estimates from the normal approximation (i.e. Equations \eqref{size0mmrm}, \eqref{size1mmrm}), 
the TS procedure and Formula \eqref{size1} are  reported for comparisons. 
In each case, $40,000$ datasets are simulated and  analyzed using MMRM  ($q=1$) with $\xv_{gi}=y_{gi0}$.
There is about $95\%$ chance that the simulated power 
 lies within $0.3\%$ of the true power.
 
We repeat the above process for a more complex MMRM. The setup is similar except that the covariates include the baseline outcome $y_{gi0}$ and a categorical prognostic factor $A$ with three levels.  We assume
that 
$$\yv_{gi} \sim N([\eta_s+\mu_{gi1}, \eta_s+ \mu_{gi2},\eta_s+\mu_{gi3}, \eta_s+\mu_{gi4}]', \Sigma)$$
for subjects in category $s$, where  $\eta_1=0$, $\eta_2=-0.5$, $\eta_3=0.5$. Each subject is in level $1$, $2$ and $3$ of the prognostic factor $A$ with probability $0.3$, $0.4$, and $0.3$ respectively.
The simulated data are analyzed using MMRM  ($q=3$) with $\xv_i=( y_{gi0}, A_{gi1}, A_{gi2})'$, where $A_{gik}=1$ if subject $i$ in group $g$ is in category $k$, and $0$ otherwise.
The   effect of  factor $A$ is assumed to  vary across visits in the analysis, but be constant over time in simulating the data.

The result is summarized in Table \ref{mmrmres}. The sample size is underestimated by the normal approximation.
The TS procedure  and Formula \eqref{size2} give similar sample size estimates. 
Formula \eqref{power0mmrm} yields power estimates that are within $0.5\%$ of the simulated power in nearly all cases.
The power equation \eqref{power1mmrm} is slightly less accurate than Equation  \eqref{power0mmrm} primarily when $\tau_4=-12$. In the worst case, the estimate by
Equation  \eqref{power1mmrm}
deviates from the simulated power by about $1.5\%$.
\end{example}

\begin{table}[htbp]
 \centering
 \def\~{\hphantom{0}}
 \begin{minipage}{150mm}
\caption{Calculated sample sizes and  power estimates  for testing $\tau_4=0$ in MMRM: \newline
$^{(a)}$ Sample size estimates are not rounded to integer values for the purpose of comparison; \newline
 $^{(b)}$ Sample size in simulation is estimated using \eqref{size2}, and rounded up to the nearest integer. The  difference in sample size between two arms is $\leq 1$;\newline
$^{(c)}$ Simulated power (SIM) based on $40,000$ simulated trials.
}\label{mmrmres}
\begin{tabular}{lccccccccccccc} \\\hline 
         &  & &  \multicolumn{5}{c}{estimated total size at $P=90\%,\alpha=0.05$ $^{(a)}$} & total &  \multicolumn{3}{c}{power ($\%$)} \\ \cline{4-8}\cline{10-12}    
          &      & inversion &   \multicolumn{2}{c}{normal} & two  & \multicolumn{2}{c}{noniterative} &size  & & \multicolumn{2}{c}{nominal}\\\cline{4-5} \cline{7-8}\cline{11-12}
 & $\tau_{41}$ & \eqref{power0mmrm}  & \eqref{size0mmrm} &  \eqref{size1mmrm} & step & \eqref{size1} &  \eqref{size2} &   $n$ $^{(b)}$ & SIM$^{(c)}$ & \eqref{power0mmrm}  & \eqref{power1mmrm} \\ \hline 
\multicolumn{11}{c}{covariates ($q=1$): baseline $\text{HAMD}_{17}$ }\\
 UN&$-12$&$ 20.31$&$ 15.24$&$ 17.16$&$ 20.85$&$ 20.03$&$ 20.44$&$ 21$&$91.33$&$91.32$&$92.36$\\  
  &$ -8$&$ 38.52$&$ 34.28$&$ 35.95$&$ 38.63$&$ 38.31$&$ 38.45$&$ 39$&$90.15$&$90.40$&$90.57$\\   
  &$ -4$&$140.88$&$137.12$&$138.67$&$140.95$&$140.83$&$140.86$&$141$&$89.92$&$90.02$&$90.02$\\   
 CS&$-12$&$ 22.48$&$ 16.77$&$ 19.29$&$ 22.78$&$ 22.22$&$ 22.60$&$ 23$&$90.37$&$90.92$&$91.78$\\  
  &$ -8$&$ 42.63$&$ 37.73$&$ 39.88$&$ 42.69$&$ 42.41$&$ 42.56$&$ 43$&$89.89$&$90.28$&$90.41$\\   
  &$ -4$&$155.37$&$150.90$&$152.89$&$155.44$&$155.31$&$155.34$&$156$&$90.15$&$90.12$&$90.12$\\   
 AR&$-12$&$ 20.53$&$ 15.34$&$ 17.34$&$ 21.06$&$ 20.25$&$ 20.67$&$ 21$&$90.89$&$90.92$&$92.05$\\  
  &$ -8$&$ 38.85$&$ 34.52$&$ 36.24$&$ 38.95$&$ 38.63$&$ 38.78$&$ 39$&$89.87$&$90.12$&$90.30$\\   
  &$ -4$&$141.91$&$138.07$&$139.67$&$141.98$&$141.85$&$141.89$&$142$&$90.18$&$90.02$&$90.01$\\   
 TOEP&$-12$&$ 18.52$&$ 13.34$&$ 15.28$&$ 19.30$&$ 18.28$&$ 18.77$&$ 19$&$91.41$&$91.09$&$92.48$\\
  &$ -8$&$ 34.28$&$ 30.02$&$ 31.66$&$ 34.41$&$ 34.04$&$ 34.21$&$ 35$&$90.65$&$90.68$&$90.90$\\   
  &$ -4$&$123.80$&$120.08$&$121.59$&$123.87$&$123.74$&$123.77$&$124$&$90.00$&$90.05$&$90.04$\\   

\vspace*{2pt}\\
\multicolumn{11}{c}{covariates ($q=3$): baseline $\text{HAMD}_{17}$, a categorical factor with three levels}\\

 UN&$-12$&$ 22.97$&$ 15.24$&$ 20.11$&$ 23.38$&$ 22.98$&$ 23.33$&$ 24$&$91.56$&$91.91$&$92.79$\\  
  &$ -8$&$ 41.07$&$ 34.28$&$ 38.52$&$ 41.14$&$ 40.89$&$ 41.03$&$ 42$&$90.56$&$90.76$&$90.92$\\   
  &$ -4$&$143.28$&$137.12$&$141.06$&$143.34$&$143.22$&$143.25$&$144$&$89.99$&$90.15$&$90.14$\\   
 CS&$-12$&$ 25.38$&$ 16.77$&$ 22.65$&$ 25.66$&$ 25.57$&$ 25.91$&$ 26$&$90.19$&$91.09$&$91.83$\\  
  &$ -8$&$ 45.44$&$ 37.73$&$ 42.75$&$ 45.50$&$ 45.30$&$ 45.45$&$ 46$&$89.93$&$90.42$&$90.55$\\   
  &$ -4$&$158.01$&$150.90$&$155.53$&$158.08$&$157.96$&$157.99$&$158$&$89.99$&$90.00$&$90.00$\\   
 AR&$-12$&$ 23.22$&$ 15.34$&$ 20.35$&$ 23.62$&$ 23.25$&$ 23.61$&$ 24$&$91.13$&$91.48$&$92.43$\\  
  &$ -8$&$ 41.44$&$ 34.52$&$ 38.85$&$ 41.51$&$ 41.26$&$ 41.40$&$ 42$&$90.21$&$90.46$&$90.64$\\   
  &$ -4$&$144.35$&$138.07$&$142.09$&$144.40$&$144.28$&$144.31$&$145$&$89.93$&$90.13$&$90.13$\\   
 TOEP&$-12$&$ 21.17$&$ 13.34$&$ 18.29$&$ 21.71$&$ 21.27$&$ 21.69$&$ 22$&$91.59$&$91.82$&$92.97$\\
  &$ -8$&$ 36.83$&$ 30.02$&$ 34.24$&$ 36.92$&$ 36.63$&$ 36.79$&$ 37$&$90.23$&$90.17$&$90.41$\\   
  &$ -4$&$126.19$&$120.08$&$123.96$&$126.24$&$126.11$&$126.15$&$127$&$90.03$&$90.19$&$90.19$\\   
  \hline
\end{tabular} 
\end{minipage}
\vspace*{6pt}
\end{table}

\section{Power and sample size for NI, equivalence and bioequivalence  trials}\label{niequiv}
The methods described in Section  \ref{genmeth} can be easily adapted for  NI, equivalence and BE trials. In these trials, inference is  made based on the confidence interval (CI) approach.
Suppose the CI for the treatment effect $\tau$ is $[c_l,c_u]= [\hat\tau -t_{f,1-\alpha/2} \sqrt{\hat{V}}, \hat\tau + t_{f,1-\alpha/2} \sqrt{\hat{V}}]$.

\subsection{Noninferiority trials}
In a NI trial, the objective is to demonstrate that the test product is not clinically  inferior to a standard treatment, or equivalently that the test treatment is not worse than the active control by 
a prespecified small amount $M_0$ called margin \cite{fda:2010, emanon:2005, hung:2007, tang:2017d}. The NI trial can be used if it would be unethical to run a placebo controlled trial or because
the new treatment may offer important advantages over the standard treatment in terms of convenience of
administration, improved safety, reduced cost, or better compliance \cite{hung:2007, tang:2017d}. 
If a lower  score indicates better health status, then $M_0>0$, and the noninferiority can be claimed when 
 the CI for $\tau$ lies below $M_0$ (i.e. $c_u<M_0$). 
If a higher score indicates better response,  noninferiority is demonstrated if the CI for $\tau$ lies above $M_0<0$ (i.e. $c_l>M_0$).
The power and sample size formulae in Section  \ref{genmeth} can be used by simply setting \cite{tang:2017d} $\tau_0=M_0$. The NI test is  one tailed, and the actual type I error is $\alpha/2$.

\subsection{Equivalence trials}\label{equivequi}
An equivalence trial aims to show that the test product is neither superior nor inferior
to the reference product, and is particularly useful in the development of biosimilar products \cite{fda:2012a}. The two treatments are not clinically different if the whole CI for $\tau$ lies completely within $[M_l,M_u]$, where $M_l<0$ and $M_u>0$ are the pre-specified lower and upper equivalence margins. 
As shown in the appendix, a generalized power formula for the equivalence test  can be obtained by extending  Phillips \cite{phillips:1990,shen:2015} approach for two sample t tests
(the true effect $\tau_1$ must lie in $[M_l,M_u]$)
\begin{equation}\label{power00equi0}
 P =\int_0^{ \frac{n(M_u-M_l)^2}{4 V t_{f,1-\alpha/2}^2}} \left[\Phi\left(\frac{M_u-\tau_1}{\sqrt{n^{-1}V}}-   t_{f,1-\frac{\alpha}{2}} \sqrt{\xi}\right) -    \Phi\left(\frac{M_l-\tau_1}{\sqrt{n^{-1}V}}+  t_{f,1-\frac{\alpha}{2}} \sqrt{\xi}\right) \right]   g(\xi) d\xi,
\end{equation}
where  $g(\xi)$ is the PDF of $\xi= \frac{\hat{V}}{{V}} \sim \frac{\chi_f^2}{f}$.
A simpler formula \cite{chow:2008, shieh:2016} that does not require numerical integration has been developed to approximate the equivalence power
\begin{equation}\label{power00equi}
 P =1-\Pr\left[t(f, \frac{M_u-\tau_1}{\sqrt{n^{-1}V}})< t_{f,1-\frac{\alpha}{2}}\right]-\Pr\left[t(f, \frac{\tau_1-M_l}{\sqrt{n^{-1}V}})< t_{f,1-\frac{\alpha}{2}}\right].
\end{equation} 
 Equation \eqref{power00equi} works very well when $n$ is large or when the estimated power is large. However, it underestimates the power, or even yields negative estimate if the sample size is too small.
The explanation is given in the Appendix.

Formula \eqref{power00equi0} is exact for the one sample t test and  two sample t test with equal variance.
Exact power  formulae for the two sample t test with unequal variance and ANCOVA with normally distributed covariates are derived in the Appendix. 


When $M_u-\tau_1=\tau_1-M_l$, the sample size formulae in Section \ref{genmeth}  can be adapted for the equivalence trial by replacing $(z_{1-\alpha/2}+z_p)^2$ and $\tau_1-\tau_0$ respectively by $(z_{1-\alpha/2}+z_{(1+P)/2})^2$  and $(M_{u}-M_{l})/2$. 
In general, there is no closed form   sample size solution in the equivalence trial. Let $\Delta_{min}=\min\{M_u-\tau_1,\tau_1-M_l\}$ and $\Delta_{max}=\max\{M_u-\tau_1,\tau_1-M_l\}$.
By the same argument as  Tang \cite{tang:2017d}, we derive the following sample size bounds based on Equation \eqref{size1}
$$ n_{g1_l} = \frac{[z_{1-\alpha/2}+z_{(1+P)/2}]^2\,V}{\Delta_{max}^2} +\frac{z_{1-\frac{\alpha}{2}}^2}{2\rho}  \leq n \leq n_{g1_u}= \frac{[z_{1-\alpha/2}+z_{(1+P)/2}]^2\,V}{\Delta_{min}^2} +\frac{z_{1-\frac{\alpha}{2}}^2}{2\rho}.$$
Similar sample size bounds can  be obtained on basis of  Equation \eqref{size2}.

\subsection{Bioequivalence trials}\label{biometh}
The purpose of the trial is to assess the BE in drug absorption between drug products \cite{chow:2001,fda:2003}, and it is useful in the development of generic drug products or 
new formulations of an existing product. The statistical principles underlying the BE and equivalence trials are the same.
In the BE trial, the PK parameters such as $C_{max}$ (maximum concentration) and AUC (area under the concentration time curve)
are used as the primary endpoints, which are approximately log-normally distributed, and 
generally log-transformed  in the analysis  \cite{chow:2001}.
Let $\mu_A^*$ and $\mu_B^*$ be the mean of the log-transformed PK parameter for product A and B respectively. 
The means of untransformed  PK parameters are $\mu_A=\exp(\mu_A^*)$ and  $\mu_B=\exp(\mu_B^*)$. 
The BE between two products can be claimed \cite{chow:2001,fda:2003} if the $90\%$ CI for  $\mu_B/\mu_A$ 
 is entirely within the BE limits of $(80\%, 125\%)$, or equivalently if the $90\%$ CI for $\mu_B^*-\mu_A^*$
lies completely within $[-0.2231,0.2231]$.
 
For drug products with relatively long half-lives, a parallel design may be used  \cite{chow:2001}.  The power and sample size formulae for the two sample t test with or without equal variances can be used directly 
by setting $\tau_1=\mu_B^*-\mu_A^*$, $M_l=-0.2231$, $M_u=0.2231$, $\alpha=0.1$, and $\sigma^2$ (or $\sigma_1^2$, $\sigma_2^2$) to be the variance of $\log(AUC)$ or $\log(C_{max})$.

A crossover design is generally preferred to reduce the sample size whenever feasible.  
The methods for  the one sample and two sample t tests  may be adapted for the crossover trial. For simplicity, we assume all subjects complete the study in the sample size calculation, and the estimated
sample size may then be adjusted  for the dropout.
Let $n_g$ be the number of subjects randomized to sequence $g$ ($g=1$ for A/B, $0$ for B/A) in a two period, two treatment crossover trial.
 Suppose the washout period is long enough so that the carryover effect is eliminated.
Let $P_{gik}$ denote the PK parameter [e.g. $\log(AUC)$] for subject $i$, period $k$, sequence $g$.
Let $d_{1i}=P_{1i2}-P_{1i1}$ for subjects in  sequence A/B, and $d_{0i}=P_{0i1}-P_{0i2}$ for sequence B/A. 
If there is no period effect, then $d_{gi} \sim N(\mu_B^*-\mu_A^*,\sigma_d^2)$ for all subjects.
The $90\%$ CI for $\mu_B^*-\mu_A^*$ is  $[\bar{d}_{..}- t_{n-1,0.95} \sqrt{\frac{\hat{\sigma}_d^2}{n}}, \bar{d}_{..} + t_{n-1,0.95} \sqrt{\frac{\hat{\sigma}_d^2}{n}}]$,  where 
$n=n_0+n_1$, $\bar{d}_{..}=\sum_{g=0}^1\sum_{i=1}^{n_g} d_{gi}/n$, and $\hat{\sigma}_d^2= \sum_{g=0}^1\sum_{i=1}^{n_i} (d_{gi}-\bar{d}_{..})^2/(n-1)$.
The methods  for the one sample t test  (described in Section \ref{equivequi}) can be used 
by setting $\tau_1=\mu_B^*-\mu_A^*$, $V=\sigma_d^2$, $f=n-1$, $\rho\approx 1$, $M_l=-0.2231$, $M_u=0.2231$ and $\alpha=0.1$.

If there is a possible period effect (denoted by $\delta$) in the crossover study, then $d_{1i}\sim N(\mu_B^*-\mu_A^*-\delta,\sigma_d^2)$, $d_{0i}\sim N(\mu_B^*-\mu_A^*+\delta,\sigma_d^2)$.
An unbiased estimate  \cite{shieh:2016} of $\mu_B^*-\mu_A^*$ is $(\bar{d}_{1}+\bar{d}_0)/2$,
and the  $90\%$ CI is  $[\frac{\bar{d}_{1}+\bar{d}_0}{2} - t_{n-2,0.95} \sqrt{\frac{n\hat{\sigma}_d^2}{4n_0n_1}}, \frac{\bar{d}_{1}+\bar{d}_0}{2}  + t_{n-2,0.95} \sqrt{\frac{n\hat{\sigma}_d^2}{4n_0n_1}}]$,  where  $\bar{d}_{g}=\sum_{i=1}^{n_g} d_{gi}/n_g$, and $\hat{\sigma}_d^2= \sum_{g=0}^1\sum_{i=1}^{n_i} (d_{gi}-\bar{d}_{g})^2/(n-2)$.
The power and sample size methods  for  the two sample t test with equal variance  (described in Section \ref{equivequi}) can be adapted
by setting $\tau_1=\mu_B^*-\mu_A^*$, $\gamma_g=n_g/n$, $\sigma^2=\sigma_d^2/4$ [i.e. $V=\sigma_d^2/(4\gamma_0\gamma_1)$], $f=n-2$, $\rho\approx 1$, $M_l=-0.2231$, $M_u=0.2231$ and $\alpha=0.1$.

In equivalence and BE trials,  the inference can be equivalently made based on the two one-sided test (TOST) procedure \cite{schuirmann:1987,phillips:1990}, and the actual type I error is $\alpha/2$.

\subsection{Numerical examples}

\begin{example}\label{bioequiexam}
\normalfont
A simulation study is conducted to assess the power and sample size  methods for a BE crossover trial. 
We set $\mu_A^*=\mu_B^*$, $\sigma^2=\sigma_d^2/4 = 0.0125k$ for $k=1,\ldots,6$.
The analysis method is described in the last paragraph in Section \ref{biometh}. 
There is no period effect ($\delta=0$) in the data simulation, but the analysis accounts for a potential period effect.  

The result is reported in Table \ref{ttestequivres}.
 The two noniterative formulae yield the sample size estimates that are the closest to the exact value. 
Formulae \eqref{power00equi0} gives the exact power estimate.
 At the target size,  formulae \eqref{power00equi} yields very accurate power approximation. However its performance deteriorates
when we reduce the required sample size by half, and the estimated power deviates from the simulated power
by about $9\%$ at $\sigma^2=\sigma_d^2/4 = 0.0125$ and $n_0=n_1=3$.

Since there is no period effect, the data can also be analyzed by the one sample t test described in Section \ref{biometh}.
The variance $\frac{\sigma_d^2}{n}$ of $\hat\mu_B^*-\hat\mu_A^*$ in the one sample t test is identical to that 
$\frac{n\hat{\sigma}_d^2}{4n_0n_1} =\frac{\sigma_d^2}{n}$ in the two sample t test when $n_0=n_1$ although the d.f. in the one sample t test is $n-1$ instead of $n-2$.
The use of the one sample t test  leads to only a minor improvement in the power, and the power estimate is presented in footnote (f) of   Table  \ref{ttestequivres}.
\end{example}

 \begin{table}[h]
\begin{center}
\caption{Calculated sample sizes and power estimates for tesing BE in a crossover trial using  the two sample t test: \newline
$^{(a)}$ Estimated using the formulae  in Sections \ref{genmeth}, where $\alpha=0.1$ and $P$ is modified as $(1+80\%)/2=0.9$ (see Section \ref{equivequi}).
The sample size estimates are not rounded  to the nearest integer for the purpose of comparison;\newline
$^{(b)}$ The exact sample size are calculated by inverting Equation \eqref{power00equi0};\newline
$^{(c)}$  The per sequence sample size  is rounded to the nearest integer;\newline
$^{(d)}$  The  per sequence sample size  is reduced by  half in order to assess the power formula \eqref{power00equi};\newline
$^{(e)}$  Simulated power (SIM) based on $1,000,000$ simulated trials. \newline
$^{(f)}$  The exact power for the one sample t test is  $79.31\%$, $78.00\%$, $81.52\%$, $80.30\%$, $79.53\%$, $80.99\%$  in the six cases.
}\label{ttestequivres}
\begin{tabular}{ccrrrc@{\extracolsep{5pt}}c@{}c@{}c@{}c@{}c@{}c@{}c@{}c@{}c} \\\hline 
    &        \multicolumn{5}{c}{estimated total  size$^{(a)}$ at $\alpha=0.1, P=80\%$} &  \\ \cline{2-6}
                        &                        &&  two & \multicolumn{2}{c}{noniterative} &    \multicolumn{4}{c}{power ($\%$)} &    \multicolumn{4}{c}{power ($\%$)}  \\\cline{5-6}\cline{7-10}  \cline{11-14}
$\sigma^2=\sigma_d^2/4$  &  exact$^{(b)}$  & normal & step & \eqref{size1} &  \eqref{size2} &  size$^{(c)}$  & SIM$^{(e)}$ & \eqref{power00equi0}$^{(f)}$ & \eqref{power00equi} &  size$^{(d)}$  & SIM$^{(e)}$ & \eqref{power00equi0} & \eqref{power00equi}  \\\hline

                                                                                    $0.0125$&$10.29$&$ 8.60$&$11.17$&$ 9.95$&$10.14$&$ 5$&$78.13$&$78.14$&$78.10$  &$ 3$&$37.88$&$37.94$&$28.74$\\
                                                                                    $0.0250$&$18.72$&$17.20$&$19.19$&$18.55$&$18.65$&$ 9$&$77.73$&$77.71$&$77.71$&$ 5$&$34.15$&$34.18$&$30.56$\\
                                                                                    $0.0375$&$27.27$&$25.80$&$27.65$&$27.15$&$27.22$&$14$&$81.38$&$81.42$&$81.42$&$ 7$&$32.11$&$32.14$&$30.13$\\
                                                                                    $0.0500$&$35.84$&$34.40$&$36.19$&$35.75$&$35.80$&$18$&$80.21$&$80.24$&$80.24$&$ 9$&$30.94$&$30.95$&$29.70$\\
                                                                                    $0.0625$&$44.42$&$43.00$&$44.75$&$44.35$&$44.39$&$22$&$79.53$&$79.49$&$79.49$&$12$&$36.67$&$36.70$&$36.41$\\
                                                                                    $0.0750$&$53.01$&$51.60$&$53.33$&$52.95$&$52.98$&$27$&$81.06$&$80.97$&$80.97$&$14$&$35.23$&$35.25$&$35.04$\\

  \hline
\end{tabular} 
\end{center}
\vspace*{6pt}
\end{table}

\begin{example}\label{mmrmexampleequi}
\normalfont
We assess the proposed methods for testing equivalence  using the two sample t test with unequal variances. We set $\tau_1=0$, $\sigma_0^2=1$, $\sigma_1^2=4$, and $\gamma_0=\gamma_1=1/2$.
For illustrative purposes, we use the margin $M_u=-M_l=0.5,1.0$ or $1.5$. Please refer to  the regulatory guidelines \cite{fda:2010,fda:2012a, emanon:2005}
on the specification of  the NI and equivalence margins.

We estimate the sample size needed to achieve $80\%$ power at $\alpha=0.05$. 
The two noniterative sample size estimates are very close to the exact size obtained by numerically inverting the power equation  \eqref{power_tsw_equi0}.
We assess the power formulae at two sample sizes. 
The exact power  by  Formula \eqref{power_tsw_equi0}  is within $0.08\%$ of the simulated power  in all cases.
 At the target  size, both formulae \eqref{powertsw_equi}  and \eqref{power00equi}  yield very good power approximations, and are much more accurate than  Equation \eqref{power00equi0}.
When we reduce the sample size by half, formulae\eqref{powertsw_equi}  and \eqref{power00equi} underestimate the power particularly at $M_u=-M_l=1.5$.
\end{example}

 \begin{table}[h]
\begin{center}
\caption{Calculated sample sizes and power estimates for tesing equivalence using two sample t tests with unequal variance: \newline
$^{(a)}$ Estimated using the formulae  in Sections \ref{genmeth}, where $\alpha=0.05$ and $P$ is modified as $(1+80\%)/2=0.9$ (see Section \ref{equivequi}).\newline
$^{(b)}$ The exact sample size are calculated by inverting Equation  \eqref{power_tsw_equi0};\newline
$^{(c)}$  The per treatment sample size  is rounded  to the nearest integer;\newline
$^{(d)}$  The  per treatment sample size  is reduced by  half in order to assess the approximate power formulae;\newline
$^{(e)}$  Simulated power (SIM) based on $1,000,000$ simulated trials. \newline
}\label{ttestunequivres}
\begin{tabular}{c@{\extracolsep{2pt}}c@{\extracolsep{5pt}}r@{\extracolsep{5pt}}r@{\extracolsep{5pt}}r@{\extracolsep{5pt}}c@{\extracolsep{2pt}}c@{\extracolsep{4pt}}c@{\extracolsep{4pt}}c@{\extracolsep{4pt}}c@{\extracolsep{4pt}}c@{\extracolsep{4pt}}c@{\extracolsep{4pt}}c@{\extracolsep{4pt}}c@{\extracolsep{4pt}}c@{\extracolsep{4pt}}c@{\extracolsep{4pt}}c@{\extracolsep{4pt}}c} \\\hline 
    &        \multicolumn{5}{c}{total  size$^{(a)}$ at $P=0.8,\alpha=0.05$} &  \\ \cline{2-6}
                        &                        &&  two & \multicolumn{2}{c}{noniterative} &    \multicolumn{6}{c}{power ($\%$)} &    \multicolumn{6}{c}{power ($\%$)}  \\\cline{5-6}\cline{7-12}  \cline{13-18}
$M_u$  &  exact$^{(b)}$  & normal & step & \eqref{size1} &  \eqref{size2} &  size$^{(c)}$  & SIM$^{(e)}$ & \eqref{power_tsw_equi0} & \eqref{powertsw_equi} & \eqref{power00equi0} & \eqref{power00equi}  &  size$^{(d)}$  & SIM$^{(e)}$ & \eqref{power_tsw_equi0} & \eqref{powertsw_equi}  & \eqref{power00equi0} & \eqref{power00equi}   \\\hline
$0.5$&$422.9$&$420.3$&$423.0$&$422.9$&$422.9$&$211$&$79.88$&$79.87$&$79.87$&$79.40$&$79.87$&$106$&$25.77$&$25.70$&$25.70$&$25.75$&$25.70$\\
$1.0$&$107.8$&$105.1$&$107.9$&$107.7$&$107.7$&$54$&$80.13$&$80.13$&$80.13$&$78.27$&$80.14$&$27$&$24.77$&$24.83$&$23.94$&$27.74$&$23.98$\\
$1.5$&$49.47$&$46.70$&$49.66$&$49.31$&$49.45$&$25$&$80.64$&$80.64$&$80.64$&$76.79$&$80.69$&$12$&$22.65$&$22.63$&$17.56$&$29.05$&$17.78$\\
  \hline
\end{tabular} 
\end{center}
\vspace*{6pt}
\end{table}

\begin{example}\label{mmrmexampleequi}
\normalfont
We assess the sample size and power determination methods for testing  equivalence  based on MMRM. The simulation setup is similar to that in Example \ref{mmrmexample}
except that the true effect is $\tau_{11}=\tau_{21}=\tau_{31}=\tau_{41}=0$. The margins satisfy $M_u=-M_l=4$ or $8$.
Since $M_u-\tau_{41}=\tau_{41}-M_l$, the noniterative sample size procedure is applicable.

The power is calculated by adapting the power equation \eqref{power0mmrm} as
\begin{eqnarray}\label{power0mmrmequi}
\begin{aligned}
P=1- \text{Pr}\left[t\left(f,\frac{M_u-\tau_{p1}}{ \sqrt{V_{\tau}^*}}\right) <  t_{f,1-\frac{\alpha}{2}}\right]- \text{Pr}\left[t\left(f,\frac{\tau_{p1}-M_l}{ \sqrt{V_{\tau}^*}}\right) <  t_{f,1-\frac{\alpha}{2}}\right].
\end{aligned}
\end{eqnarray}

Table \ref{mmrmresequi} summarizes the results, and the performance is comparable to that for superiority tests reported in Example \ref{mmrmexample}.
\end{example}

\begin{table}[htbp]
 \centering
 \def\~{\hphantom{0}}
 \begin{minipage}{150mm}
\caption{Calculated sample sizes and  power estimates  for testing equivalence at visit $p$ in MMRM: \newline 
$^{(a)}$ Sample size estimates are not rounded to integer values for the purpose of comparison; \newline
 $^{(b)}$ Sample size in simulation is estimated via \eqref{size2} [$P$ is modified as $(1+90\%)/2=0.95$], and rounded up to the nearest integer. The  difference in sample size between two arms is $\leq 1$; \newline
$^{(c)}$ Simulated power (SIM) based on $40,000$ simulated trials.
}\label{mmrmresequi}
\begin{tabular}{lccccccccccccccc} \\\hline 
         &  & &  \multicolumn{5}{c}{estimated total size at $P=90\%,\alpha=0.05$ $^{(a)}$} & total &  \\ \cline{4-8}
          &Margin      & inversion &   \multicolumn{2}{c}{normal} & two  & \multicolumn{2}{c}{noniterative} &size  &   \multicolumn{2}{c}{power ($\%$)}\\\cline{4-5} \cline{7-8}\cline{10-11}
 & $M_u=-M_l$ & \eqref{power0mmrmequi}  & \eqref{size0mmrm} &  \eqref{size1mmrm} & step & \eqref{size1} &  \eqref{size2} &   $n$ $^{(b)}$ & SIM$^{(c)}$ & \eqref{power0mmrmequi}  \\ \hline 
\multicolumn{11}{c}{covariates ($q=1$): baseline $\text{HAMD}_{17}$ }\\

 UN &$ 8$&$ 46.52$&$ 42.40$&$ 44.03$&$ 46.83$&$ 46.33$&$ 46.45$&$ 47$&$90.60$&$90.42$\\   
  &$ 4$&$173.32$&$169.58$&$171.12$&$173.58$&$173.27$&$173.29$&$174$&$90.22$&$90.15$\\   
 CS  &$ 8$&$ 51.45$&$ 46.66$&$ 48.77$&$ 51.75$&$ 51.26$&$ 51.38$&$ 52$&$89.83$&$90.43$\\   
  &$ 4$&$191.06$&$186.62$&$188.60$&$191.36$&$191.01$&$191.04$&$192$&$90.05$&$90.19$\\   
 AR  &$ 8$&$ 46.90$&$ 42.69$&$ 44.38$&$ 47.21$&$ 46.71$&$ 46.83$&$ 47$&$90.19$&$90.09$\\   
  &$ 4$&$174.57$&$170.75$&$172.34$&$174.84$&$174.52$&$174.55$&$175$&$89.99$&$90.09$\\   
 TOEP  &$ 8$&$ 41.25$&$ 37.13$&$ 38.73$&$ 41.58$&$ 41.05$&$ 41.18$&$ 42$&$90.66$&$90.73$\\   
  &$ 4$&$152.20$&$148.51$&$150.01$&$152.46$&$152.14$&$152.17$&$153$&$90.26$&$90.20$\\
\vspace*{2pt}\\
\multicolumn{11}{c}{covariates ($q=3$): baseline $\text{HAMD}_{17}$, a categorical factor with three levels}\\
 UN &$ 8$&$ 49.04$&$ 42.40$&$ 46.55$&$ 49.31$&$ 48.86$&$ 48.97$&$ 49$&$89.57$&$89.97$\\   
  &$ 4$&$175.70$&$169.58$&$173.50$&$175.96$&$175.65$&$175.68$&$176$&$89.97$&$90.06$\\   
 CS&$ 8$&$ 54.22$&$ 46.66$&$ 51.57$&$ 54.52$&$ 54.09$&$ 54.20$&$ 55$&$89.68$&$90.61$\\   
  &$ 4$&$193.70$&$186.62$&$191.23$&$193.99$&$193.65$&$193.68$&$194$&$89.81$&$90.06$\\   
 AR  &$ 8$&$ 49.46$&$ 42.69$&$ 46.94$&$ 49.73$&$ 49.28$&$ 49.39$&$ 50$&$90.14$&$90.47$\\   
  &$ 4$&$176.99$&$170.75$&$174.76$&$177.25$&$176.93$&$176.96$&$177$&$89.76$&$90.00$\\   
 TOEP &$ 8$&$ 43.77$&$ 37.13$&$ 41.25$&$ 44.06$&$ 43.58$&$ 43.70$&$ 44$&$90.06$&$90.24$\\   
  &$ 4$&$154.57$&$148.51$&$152.37$&$154.82$&$154.51$&$154.54$&$155$&$90.00$&$90.11$\\  
  \hline
\end{tabular} 
\end{minipage}
\vspace*{6pt}
\end{table}

Simulation also demonstrates the accuracy of the power and sample size formulae for ANCOVA in equivalence trials. 
The results are not reported due to limited space.
Sample  SAS codes for the power and sample size determinations for t tests, ANCOVA and MMRM in superiority, NI and equivalence trials are provided in the Supporting Information.

 \section{Discussion}
We develop a generalized sample size  procedure for  t tests by modifying and extending 
Guenther's method for the one sample and two sample t tests. The procedure is simple and noniterative by adding a few correction terms
to the sample size from the normal approximation. Numerical examples demonstrate its excellent performance.
Both formulae \eqref{size1} and \eqref{size2} slightly outperform the TS procedure, and are much more accurate than the approaches based on the normal 
approximation or the asymptotic variance in small and moderate samples. 

Formula \eqref{size2} tends to be slightly more accurate than formula \eqref{size1} for the one-sample and two sample t tests.
In ANCOVA and MMRM, the  noniterative procedure (particularly formula \eqref{size2}) has a tendency to slightly overestimate the required size
 if the number of covariates is relatively large, and the total
 size is small (possibly because of the approximation method used to handle the covariates).  However, these scenarios  rarely happen in practice.
 Let's take the last case in Table \ref{ancovares} as an example. In this case, $q=3, n\approx 14$, and the total number of model parameters is $6$.
If the model includes too many covariates, the power may actually reduce, and
the parameter estimate may not be consistent \cite{kahan:2014}. 
The regulatory guideline \cite{chmp:2013} recommends that the primary analysis  shall include only a few important covariates. 

Since the final sample size takes only integer values, the estimate from the noniterative procedure after rounding is generally exact or nearly exact (deviate from the target sample size by at most $1$ in our examples).
It would be beneficial to evaluate the power at several integer sample sizes near the noniterative estimate in order to find the most appropriate sample size.
It is a common practice to round the total sample size or the size per treatment arm up to the next integer, and it ensures that the actual power is at least as large as the target power.
A smaller sample size may also be used sometimes. For example, in case $5$ ($\mu_1-\mu_0=1.5$ and the exact size is $16.12$) in Table \ref{ttestres}, 
we may round the total sample size down to $n=16$ if it is extremely difficult to enroll patients (e.g. in rare disease trials) since the exact power $79.65\%$ at $n=16$ is almost close to
 the target $80\%$ power.

An extensive literature \cite{boos:2000,sullivan:2003, tang:2017} indicates that  the t tests,
 ANCOVA and MMRM are fairly robust to deviations from non-normality.  
As confirmed by unreported simulation studies (see also Tang \cite{tang:2017}), the proposed sample size procedure works well for mild to moderate nonnormal data.
It is always recommended to 
verify the power and sample size estimate by simulations particularly when the data are non-normal or the sample size is small.
We have focused on the unstratified trials. In a companion paper, we will investigate the power and sample size determination for testing the main treatment effect and 
treatment $\times$ stratum interaction in stratified trials using ANCOVA \cite{tang:2018e}.

\appendix
\section*{Appendix: Technical Proofs}
\begin{proof}{\bf of equations \eqref{size1} and \eqref{size2}:} 
The type I error and power are calculated by assuming that $Z_1=[\sqrt{n}(\hat\tau -\tau)- c (\sqrt{\hat{V}}- \sqrt{V})]/\sigma_z \sim N(0,1)$
and $Z_2=[\sqrt{n}(\hat\tau -\tau)+ c (\sqrt{\hat{V}}- \sqrt{V})]/\sigma_z \sim N(0,1)$,
where $c=t_{f, 1-\alpha/2}$, $\sigma_z^2=V[1+c^2 /(2f)]$, $\tau=\tau_0$ under $H_0$, and $\tau=\tau_1$ under $H_1$.
The type I error of the test  ($\tau=\tau_0$ under $H_0$) is
\begin{eqnarray*}\label{cdeter0}
\begin{aligned}
&\Pr(|T|>t_{f,1-\frac{\alpha}{2}})=\Pr\left[ Z_1  > c\sqrt{\frac{V}{\sigma_z^2}} \,\right]+ \Pr\left[Z_2  <-c\sqrt{\frac{V}{\sigma_z^2}} \,\right] = 2\Phi\left[-c\sqrt{\frac{V}{\sigma_z^2}}\,\right].
\end{aligned}
\end{eqnarray*}
Setting the type I error at $\alpha$ yields an approximation of the critical value $c=t_{f, 1-\alpha/2}$
\begin{equation*}\label{cdeter1}
c\sqrt{\frac{V}{\sigma_z^2}}  = z_{1-\frac{\alpha}{2}}  \text { and }   c =  z_{1-\frac{\alpha}{2}} \sqrt{\frac{2f}{2f - z_{1-\frac{\alpha}{2}}^2}}.                           
\end{equation*}
The power  ($\tau=\tau_1$ under $H_1$) is approximately
\begin{eqnarray*}\label{power1}
\begin{aligned}
&P= \Pr(|T|>t_{f,1-\frac{\alpha}{2}}) =\Phi\left[-Z_1  <  \frac{\tau_1-\tau_0}{\sqrt{n^{-1}\sigma_z^2}} - c\sqrt{\frac{V}{\sigma_z^2}}\, \right] 
+\Phi\left[Z_2  <  \frac{\tau_0-\tau_1}{\sqrt{n^{-1}\sigma_z^2}} - c\sqrt{\frac{V}{\sigma_z^2}}\,\right] \approx \Phi\left[\frac{|\tau_1-\tau_0|}{\sqrt{n^{-1}\sigma_z^2}} -z_{1-\frac{\alpha}{2}}\right].
\end{aligned}
\end{eqnarray*}

Inverting the above power formula  yields the sample  size
\begin{eqnarray*}
\begin{aligned}
n_{\text{g}1}= \frac{(z_{1-\frac{\alpha}{2}}+z_P)^2\,\sigma_z^2}{(\tau_1-\tau_0)^2} \approx  \tilde{n}+h_f \frac{z_{1-\frac{\alpha}{2}}^2}{2\rho},
\end{aligned}
\end{eqnarray*}
where $\rho=f/n\approx f/\tilde{n}$ and $h_f= 2f/(2f-z_{1-\alpha/2}^2)$. Equations \eqref{size1} and \eqref{size2} are obtained respectively by approximating $h_f \approx 1$, and
 $h_f\approx 1+  z_{1-\alpha/2}^2/(2n_{\text{g}1}\rho)$.
\end{proof}

\flushleft{{\bf Solution of Equation \eqref{sizelowunstra}:}}
 Equation \eqref{sizelowunstra} can be reorganized as $\tilde{n}^2 -\tilde{n}(n_{\text{asy}}+q+3)+3n_{\text{asy}}=0$. Its solution is
$$ \tilde{n}= \frac{(n_{\text{asy}}+q+3) +\sqrt{(n_{\text{asy}}+q+3)^2-12n_{\text{asy}} }}{2}.$$
A little algebra shows that $n_{\text{asy}}+q-3 < \sqrt{(n_{\text{asy}}+q+3)^2-12n_{\text{asy}} }< n_{\text{asy}}+q+3$ by noting that $q>0$ is a positive integer. Thus
$n_{\text{asy}}+q< \tilde{n}< n_{\text{asy}}+q+3$.

\begin{proof}{\bf of Equation \eqref{mmrmdf}}:
By Tang \cite{tang:2017}, we have  $\widehat{\text{var}}(\hat\betav_j)=\hat\sigma_j^2 (Y_{j}'Q_{j} Y_{j})^{-1}$,
 $\widehat{\text{var}}(\hat\sigma_j^2)= 2\hat\sigma_j^4/(m_j-q^*)$, $\partial l_{pj}/\partial\betav_k = l_{pk}(l_{1j},\ldots,l_{k-1,j})'$ [it is $\zerov$ if $j\geq k$].
Thus $\partial (\sum_{j=1}^p \hat{l}_{pj}^2\hat\sigma_j^2 V_{x_j})/\partial\betav_k =2 l_{pk} \av_k L_{k-1}'$.
 Also $\hat\betav_j$'s and $\hat\sigma_j^2$'s are independent. 
Equation \eqref{mmrmdf} holds since  by delta method, we have
$$\widehat{\text{var}}(\sum_{j=1}^p \hat{l}_{pj}^2\hat\sigma_j^2 V_{x_j})
=4\sum_{j=2}^pA_j + 2\sum_{j=1}^p \frac{ \hat{l}_{pj}^2 a_j^2 }{m_j-q^*}.$$
\end{proof}

{\flushleft{\bf Derivation of $\omega_{jt}$'s in Equation \eqref{vartau}}:}
The variance of $\hat\betav$ is given by $$ \text{var}(\hat\betav_j) = \sigma_j^2 \text{E}[ (Y_{j}'Q_{j} Y_{j})^{-1}] = \frac{\sigma_j^2}{m_j-q^*-j} \Sigma_{j-1}^{-1}$$
since $Y_{j}'Q_{j} Y_{j}$ follows a Wishart distribution with  $m_j-q^*$ d.f. and scale matrix $\Sigma_{j-1}$, where 
$\Sigma_{j-1}$ is the leading $(j-1)\times (j-1)$ submatrix of $\Sigma$. Thus $L_{j-1}' \text{\normalfont var}(\hat\betav_j)  L_{j-1}$ is a diagnoal matrix, and its 
 $(t,t)$ entry is given by $\omega_{jt}=\sigma_j^2/[(m_j-q^*-j)\sigma_t^2]$.

{\flushleft{\bf Derivation of $c_j$'s in Equation \eqref{normfrho}}:}
By Tang \cite{tang:2017}, $\text{E}(\hat{l}^2_{pj})\sigma_j^2 =l_{pj}^2\sigma_j^2 +\text{var}(\hat{l}_{pj})\sigma_j^2 =l_{pj}^2\sigma_j^2 +\sum_{k=j+1}^pl_{pk}^2 \omega_{kj}\sigma_j^2 $, and
$c_j= \text{E}(\hat{ l}_{pj}^2\hat\sigma_j^2) = (1-\frac{j-1}{m_j-q^*})\text{E}(\hat{l}^2_{pj})\sigma_j^2$, where
$\omega_{kj}\sigma_j^2 = \sigma_k^2/(m_k-q^*-k)$. 

\begin{proof}{\bf of Equations \eqref{power00equi0} and \eqref{power00equi}}: Let $\xi = \frac{\hat{V}}{V}$. Then $f\xi \sim \chi_f^2$. 
Conditioning on $\hat{V}$,  the equivalence power is given by
\begin{eqnarray*}
\begin{aligned}
\varphi(\xi) = \Pr(c_l>M_l, c_u<M_u|\hat{V}) = \Phi(\frac{M_u-\tau_1}{\sqrt{n^{-1}V}} -t_{f,1-\frac{\alpha}{2}} \sqrt{\xi})
        -\Phi(\frac{M_l-\tau_1}{\sqrt{n^{-1}V}}+ t_{f,1-\frac{\alpha}{2}} \sqrt{\xi})
\end{aligned}
\end{eqnarray*}
if $\xi$ lies in the region $A_\xi=\left\{\xi: \frac{M_u-M_l}{\sqrt{n^{-1}V}} < 2 t_{f,1-\frac{\alpha}{2}} \sqrt{\xi}\right\}$, and $0$ otherwise since $\varphi(\xi)<0$ when $\xi\notin A_\xi$.
The power is $\int  \varphi(\xi) I(\xi \in A_\xi) g(\xi)d\xi$, and this leads to Equation \eqref{power00equi0}. 

Equation \eqref{power00equi} is obtained as $P\approx \int  \varphi(\xi)  g(\xi)d\xi$, and the  approximation error $\int  \varphi(\xi) I(\xi \notin A_\xi) g(\xi)d\xi$ is negative.
 When $n$ is small, there is a large chance that $\xi\notin A_\xi$, leading to a large error in the power estimation.
\end{proof}

{\flushleft{\bf Exact power formula for testing equivalence using ANCOVA with normally distributed covariates:}\\
By using the same argument as that for Equation \eqref{power00equi0}, we get the exact power equation 
\begin{eqnarray}\label{power_ancova_equi0}
\begin{aligned}
P = \int_0^\infty \int_0^{\frac{(M_u-M_l)^2}{4\sigma^2V_x(\tilde{\Upsilon})  t_{f,1-\alpha/2}^2 } }  \left[ \Phi(\frac{M_u-\tau_1}{\sqrt{\sigma^2 V_x(\tilde{\Upsilon})}} -t_{f,1-\frac{\alpha}{2}} \sqrt{\xi})
        -\Phi(\frac{M_l-\tau_1}{\sqrt{\sigma^2 V_x(\tilde{\Upsilon})}}+ t_{f,1-\frac{\alpha}{2}} \sqrt{\xi})\right] g(\xi)  g(\tilde{\Upsilon})d\xi d \tilde{\Upsilon},
\end{aligned}
\end{eqnarray}
where $g(\xi)$ is the PDF of $\xi=\hat\sigma^2/\sigma^2\sim \chi_f^2/f$. In large samples, Equation \eqref{power_ancova_equi0} can be well approximated by
\begin{eqnarray}\label{power_ancova_equi}
\begin{aligned}
P =1-\int \text{Pr}\left[ t\left(f,\frac{(M_u-\tau_1)}{\sqrt{\sigma^2V_x(\tilde{\Upsilon})}}\,\right) < t_{f,1-\frac{\alpha}{2}} \right] g(\tilde{\Upsilon})d \tilde{\Upsilon}
-\int \text{Pr}\left[ t\left(f,\frac{(\tau_1-M_l)}{\sqrt{\sigma^2V_x(\tilde{\Upsilon})}}\,\right) < t_{f,1-\frac{\alpha}{2}} \right] g(\tilde{\Upsilon}) d \tilde{\Upsilon}.
\end{aligned}
\end{eqnarray}

{\flushleft{\bf Exact power formula for testing equivalence using two sample t test with unequal variances:}\\
We extend Moser {\it et al} \cite{moser:1989} exact approach to equivalence trials. 
Note that $u=s_1^2\sigma_0^2/(s_0^2\sigma_1^2)\sim F(n_1-1,n_0-1)$ is independent of $\xi= \frac{(n_1-1)\frac{s_1^2}{\sigma_1^2}+ (n_0-1)\frac{s_0^2}{\sigma_0^2}}{n-2} \sim \frac{\chi_{n-2}^2}{n-2}$.
Let 
 $$V(u) = \frac{(n-2)}{(n_1-1)u+(n_0-1)}[\frac{u\sigma_1^2}{n_1}+\frac{\sigma_0^2}{n_0}], \,\,
 f(u)= \frac{\left[\frac{u\sigma_1^2}{n_1} +\frac{\sigma_0^2}{n_0}\right]^2 }{\frac{u^2\sigma_1^4}{n_1^2(n_1-1)} +\frac{\sigma_0^4}{n_0^2(n_0-1)} }, \text{ and }
h(u)=t_{f(u),1-\frac{\alpha}{2}} \sqrt{ \frac{V(u)}{\frac{\sigma_1^2}{n_1}+ \frac{\sigma_0^2}{n_0}}}.$$
Then $\frac{s_1^2}{n_1}+\frac{s_0^2}{n_0}=\xi\,V(u)$.
Let $g(u)$ and $g(\xi)$ denote respectively the PDF of $u$ and $\xi$. The exact power is given by
\begin{eqnarray}\label{power_tsw_equi0}
\begin{aligned}
P = \int_0^\infty \int_0^{c(u)}  \left[ \Phi\left(\frac{M_u-\tau_1}{\sqrt{\frac{\sigma_1^2}{n_1}+\frac{\sigma_0^2}{n_0} }} -h(u) \sqrt{\xi} \right)
      -  \Phi\left(\frac{M_l-\tau_1}{\sqrt{\frac{\sigma_1^2}{n_1}+\frac{\sigma_0^2}{n_0} }}+  h(u) \sqrt{\xi }\right)\right] g(\xi) g(u)d\xi d u.
\end{aligned}
\end{eqnarray}
where $c(u)=\frac{(M_u-M_l)^2}{ 4V(u)\,t_{f(u),1-\alpha/2}^2 }$.
At a large sample size, the power can be approximated by
\begin{eqnarray}\label{powertsw_equi}
\begin{aligned}
P =1-\int\left\{ \text{Pr}\left[ t\left(n-2,\frac{M_u-\tau_1}{ \sqrt{\frac{\sigma_1^2}{n_1}+\frac{\sigma_0^2}{n_0}}}\,\right) < h(u) \right]
-\text{Pr}\left[ t\left(n-2,\frac{\tau_1-M_l}{\sqrt{ \frac{\sigma_1^2}{n_1}+\frac{\sigma_0^2}{n_0}}}\,\right) <h(u) \right]\right\}g(u)du.
\end{aligned}
\end{eqnarray}
Setting $(M_l, M_u)=(-\infty,\tau_0)$ or $(\tau_0,\infty)$ into \eqref{powertsw_equi} yields the  formula obtained by Moser {\it et al} \cite{moser:1989}, which is suitable for superiority and NI tests.

\bibliographystyle{wileyj}
\bibliography{tsize} 

\end{document}